\documentclass[aps,prd,twocolumn,reprint,showpacs,superscriptaddress,floatfix,amsmath,amssymb,nopreprintnumbers]{revtex4-1}

\usepackage{natbib}
\usepackage{times}
\usepackage{mathptm}
\usepackage{graphicx}
\usepackage{wrapfig}
\usepackage{subfigure}
\usepackage{morefloats}
\usepackage{hyperref}
\usepackage{scrextend}

\newcommand{\figsize}{0.495\textwidth} 
\newcommand{\figsizestar}{\textwidth} 

\newcommand{\ppbar}{\ensuremath{p\bar{p}}}

\newcommand{\Zg} {\mbox{\ensuremath{Z/\gamma^*}}}
\newcommand{\Z}  {\ensuremath{Z}}
\newcommand{\Zll}    {\mbox{\ensuremath{Z/\gamma^* \rightarrow \ell^+\ell^-}}}
\newcommand{\Zee}    {\mbox{\ensuremath{Z/\gamma^* \rightarrow e^+e^-}}}
\newcommand{\Zmm}    {\mbox{\ensuremath{Z/\gamma^* \rightarrow \mu^+\mu^-}}}
\newcommand{\Ztt}    {\mbox{\ensuremath{Z/\gamma^* \rightarrow \tau^+\tau^-}}}

\newcommand{\Zjets}{\mbox{\ensuremath{Z/\gamma^* + \textrm{jets}}}}
\newcommand{\Zonejet}{\mbox{\ensuremath{Z/\gamma^* + \geqslant 1~\textrm{jet}}}}
\newcommand{\Ztwojets}{\mbox{\ensuremath{Z/\gamma^* + \geqslant 2~\textrm{jets}}}}
\newcommand{\Zthreejets}{\mbox{\ensuremath{Z/\gamma^* + \geqslant 3~\textrm{jets}}}}
\newcommand{\Zfourjets}{\mbox{\ensuremath{Z/\gamma^* + \geqslant 4~\textrm{jets}}}}
\newcommand{\ZNjets}{\mbox{\ensuremath{Z/\gamma^* + \geqslant N_{\textrm{jets}}}}}

\newcommand{\Zeejets}{\ensuremath{Z/\gamma^* \rightarrow e^+e^- + \textrm{jets}}}
\newcommand{\Zmmjets}{\ensuremath{Z/\gamma^* \rightarrow \mu^+\mu^- + \textrm{jets}}}
\newcommand{\Zttjets}{\ensuremath{Z/\gamma^* \rightarrow \tau^+\tau^- + \textrm{jets}}}

\newcommand{\ZeeNjets}{\ensuremath{Z/\gamma^* \rightarrow e^+e^- + \geqslant  N_{\textrm{jets}}}}
\newcommand{\ZmmNjets}{\ensuremath{Z/\gamma^* \rightarrow \mu^+\mu^- + \geqslant  N_{\textrm{jets}}}}

\newcommand{\ttbar}{\mbox{\ensuremath{t \overline{t}}}}

\newcommand{\Wjets}{\mbox{\ensuremath{W + \textrm{jets}}}}
\newcommand{\W}{\ensuremath{W}}

\newcommand{\pt}{\ensuremath{p_{\textrm{T}}}}
\newcommand{\et}{\ensuremath{E_{\textrm{T}}}}
\newcommand{\htjet}{\ensuremath{H_{\textrm{T}}^{\textrm{jet}}}}

\newcommand{\ptjet}   {\ensuremath{p_{\textrm{T}}^{\textrm{jet}}}}
\newcommand{\ptcal}   {\ensuremath{p_{\textrm{T,cal}}}}
\newcommand{\yjet}    {\ensuremath{y^{\textrm{jet}}}}

\newcommand{\pythia} {{\sc pythia}}
\newcommand{\blackhatsherpa} {{\sc blackhat+sherpa}}
\newcommand{\blackhat} {{\sc blackhat}}

\newcommand{\alpgen} {{\sc alpgen}}
\newcommand{\alpgenpythia} {{\sc alpgen+pythia}}
\newcommand{\powheg} {{\sc powheg}}
\newcommand{\powhegpythia} {{\sc powheg+pythia}}
\newcommand{\loopsim} {{\sc loopsim}}
\newcommand{\loopsimmcfm} {{\sc loopsim+mcfm}}
\newcommand{\mcfm} {{\sc mcfm}}
\newcommand{\nnlo} {$\overline{\textrm{n}}$NLO}
\newcommand{\nloqcdew} {{\sc nlo qcd $\otimes$ nlo ew}}

\newcommand{\gevc} {GeV/$c$}
\newcommand{\gevcsq} {GeV/$c^2$}

\begin{document}

\title{Measurement of differential production cross section for \Zg{} bosons
in association with jets in \ppbar{} collisions at $\sqrt{s}=1.96$~TeV \\ \vspace*{2.0cm}}

\affiliation{Institute of Physics, Academia Sinica, Taipei, Taiwan 11529, Republic of China}
\affiliation{Argonne National Laboratory, Argonne, Illinois 60439, USA}
\affiliation{University of Athens, 157 71 Athens, Greece}
\affiliation{Institut de Fisica d'Altes Energies, ICREA, Universitat Autonoma de Barcelona, E-08193, Bellaterra (Barcelona), Spain}
\affiliation{Baylor University, Waco, Texas 76798, USA}
\affiliation{Istituto Nazionale di Fisica Nucleare Bologna, \ensuremath{^{ii}}University of Bologna, I-40127 Bologna, Italy}
\affiliation{University of California, Davis, Davis, California 95616, USA}
\affiliation{University of California, Los Angeles, Los Angeles, California 90024, USA}
\affiliation{Instituto de Fisica de Cantabria, CSIC-University of Cantabria, 39005 Santander, Spain}
\affiliation{Carnegie Mellon University, Pittsburgh, Pennsylvania 15213, USA}
\affiliation{Enrico Fermi Institute, University of Chicago, Chicago, Illinois 60637, USA}
\affiliation{Comenius University, 842 48 Bratislava, Slovakia; Institute of Experimental Physics, 040 01 Kosice, Slovakia}
\affiliation{Joint Institute for Nuclear Research, RU-141980 Dubna, Russia}
\affiliation{Duke University, Durham, North Carolina 27708, USA}
\affiliation{Fermi National Accelerator Laboratory, Batavia, Illinois 60510, USA}
\affiliation{University of Florida, Gainesville, Florida 32611, USA}
\affiliation{Laboratori Nazionali di Frascati, Istituto Nazionale di Fisica Nucleare, I-00044 Frascati, Italy}
\affiliation{University of Geneva, CH-1211 Geneva 4, Switzerland}
\affiliation{Glasgow University, Glasgow G12 8QQ, United Kingdom}
\affiliation{Harvard University, Cambridge, Massachusetts 02138, USA}
\affiliation{Division of High Energy Physics, Department of Physics, University of Helsinki, FIN-00014, Helsinki, Finland; Helsinki Institute of Physics, FIN-00014, Helsinki, Finland}
\affiliation{University of Illinois, Urbana, Illinois 61801, USA}
\affiliation{The Johns Hopkins University, Baltimore, Maryland 21218, USA}
\affiliation{Institut f\"{u}r Experimentelle Kernphysik, Karlsruhe Institute of Technology, D-76131 Karlsruhe, Germany}
\affiliation{Center for High Energy Physics: Kyungpook National University, Daegu 702-701, Korea; Seoul National University, Seoul 151-742, Korea; Sungkyunkwan University, Suwon 440-746, Korea; Korea Institute of Science and Technology Information, Daejeon 305-806, Korea; Chonnam National University, Gwangju 500-757, Korea; Chonbuk National University, Jeonju 561-756, Korea; Ewha Womans University, Seoul, 120-750, Korea}
\affiliation{Ernest Orlando Lawrence Berkeley National Laboratory, Berkeley, California 94720, USA}
\affiliation{University of Liverpool, Liverpool L69 7ZE, United Kingdom}
\affiliation{University College London, London WC1E 6BT, United Kingdom}
\affiliation{Centro de Investigaciones Energeticas Medioambientales y Tecnologicas, E-28040 Madrid, Spain}
\affiliation{Massachusetts Institute of Technology, Cambridge, Massachusetts 02139, USA}
\affiliation{University of Michigan, Ann Arbor, Michigan 48109, USA}
\affiliation{Michigan State University, East Lansing, Michigan 48824, USA}
\affiliation{Institution for Theoretical and Experimental Physics, ITEP, Moscow 117259, Russia}
\affiliation{University of New Mexico, Albuquerque, New Mexico 87131, USA}
\affiliation{The Ohio State University, Columbus, Ohio 43210, USA}
\affiliation{Okayama University, Okayama 700-8530, Japan}
\affiliation{Osaka City University, Osaka 558-8585, Japan}
\affiliation{University of Oxford, Oxford OX1 3RH, United Kingdom}
\affiliation{Istituto Nazionale di Fisica Nucleare, Sezione di Padova, \ensuremath{^{jj}}University of Padova, I-35131 Padova, Italy}
\affiliation{University of Pennsylvania, Philadelphia, Pennsylvania 19104, USA}
\affiliation{Istituto Nazionale di Fisica Nucleare Pisa, \ensuremath{^{kk}}University of Pisa, \ensuremath{^{ll}}University of Siena, \ensuremath{^{mm}}Scuola Normale Superiore, I-56127 Pisa, Italy, \ensuremath{^{nn}}INFN Pavia, I-27100 Pavia, Italy, \ensuremath{^{oo}}University of Pavia, I-27100 Pavia, Italy}
\affiliation{University of Pittsburgh, Pittsburgh, Pennsylvania 15260, USA}
\affiliation{Purdue University, West Lafayette, Indiana 47907, USA}
\affiliation{University of Rochester, Rochester, New York 14627, USA}
\affiliation{The Rockefeller University, New York, New York 10065, USA}
\affiliation{Istituto Nazionale di Fisica Nucleare, Sezione di Roma 1, \ensuremath{^{pp}}Sapienza Universit\`{a} di Roma, I-00185 Roma, Italy}
\affiliation{Mitchell Institute for Fundamental Physics and Astronomy, Texas A\&M University, College Station, Texas 77843, USA}
\affiliation{Istituto Nazionale di Fisica Nucleare Trieste, \ensuremath{^{qq}}Gruppo Collegato di Udine, \ensuremath{^{rr}}University of Udine, I-33100 Udine, Italy, \ensuremath{^{ss}}University of Trieste, I-34127 Trieste, Italy}
\affiliation{University of Tsukuba, Tsukuba, Ibaraki 305, Japan}
\affiliation{Tufts University, Medford, Massachusetts 02155, USA}
\affiliation{University of Virginia, Charlottesville, Virginia 22906, USA}
\affiliation{Waseda University, Tokyo 169, Japan}
\affiliation{Wayne State University, Detroit, Michigan 48201, USA}
\affiliation{University of Wisconsin, Madison, Wisconsin 53706, USA}
\affiliation{Yale University, New Haven, Connecticut 06520, USA}

\author{T.~Aaltonen}
\affiliation{Division of High Energy Physics, Department of Physics, University of Helsinki, FIN-00014, Helsinki, Finland; Helsinki Institute of Physics, FIN-00014, Helsinki, Finland}
\author{S.~Amerio\ensuremath{^{jj}}}
\affiliation{Istituto Nazionale di Fisica Nucleare, Sezione di Padova, \ensuremath{^{jj}}University of Padova, I-35131 Padova, Italy}
\author{D.~Amidei}
\affiliation{University of Michigan, Ann Arbor, Michigan 48109, USA}
\author{A.~Anastassov\ensuremath{^{v}}}
\affiliation{Fermi National Accelerator Laboratory, Batavia, Illinois 60510, USA}
\author{A.~Annovi}
\affiliation{Laboratori Nazionali di Frascati, Istituto Nazionale di Fisica Nucleare, I-00044 Frascati, Italy}
\author{J.~Antos}
\affiliation{Comenius University, 842 48 Bratislava, Slovakia; Institute of Experimental Physics, 040 01 Kosice, Slovakia}
\author{G.~Apollinari}
\affiliation{Fermi National Accelerator Laboratory, Batavia, Illinois 60510, USA}
\author{J.A.~Appel}
\affiliation{Fermi National Accelerator Laboratory, Batavia, Illinois 60510, USA}
\author{T.~Arisawa}
\affiliation{Waseda University, Tokyo 169, Japan}
\author{A.~Artikov}
\affiliation{Joint Institute for Nuclear Research, RU-141980 Dubna, Russia}
% ...

\collaboration{CDF Collaboration}
\altaffiliation[With visitors from]{
\ensuremath{^{a}}University of British Columbia, Vancouver, BC V6T 1Z1, Canada,
\ensuremath{^{b}}Istituto Nazionale di Fisica Nucleare, Sezione di Cagliari, 09042 Monserrato (Cagliari), Italy,
\ensuremath{^{c}}University of California Irvine, Irvine, CA 92697, USA,
\ensuremath{^{d}}Institute of Physics, Academy of Sciences of the Czech Republic, 182~21, Czech Republic,
\ensuremath{^{e}}CERN, CH-1211 Geneva, Switzerland,
\ensuremath{^{f}}Cornell University, Ithaca, NY 14853, USA,
\ensuremath{^{g}}University of Cyprus, Nicosia CY-1678, Cyprus,
\ensuremath{^{h}}Office of Science, U.S. Department of Energy, Washington, DC 20585, USA,
\ensuremath{^{i}}University College Dublin, Dublin 4, Ireland,
\ensuremath{^{j}}ETH, 8092 Z\"{u}rich, Switzerland,
\ensuremath{^{k}}University of Fukui, Fukui City, Fukui Prefecture, Japan 910-0017,
\ensuremath{^{l}}Universidad Iberoamericana, Lomas de Santa Fe, M\'{e}xico, C.P. 01219, Distrito Federal,
\ensuremath{^{m}}University of Iowa, Iowa City, IA 52242, USA,
\ensuremath{^{n}}Kinki University, Higashi-Osaka City, Japan 577-8502,
\ensuremath{^{o}}Kansas State University, Manhattan, KS 66506, USA,
\ensuremath{^{p}}Brookhaven National Laboratory, Upton, NY 11973, USA,
\ensuremath{^{q}}Queen Mary, University of London, London, E1 4NS, United Kingdom,
\ensuremath{^{r}}University of Melbourne, Victoria 3010, Australia,
\ensuremath{^{s}}Muons, Inc., Batavia, IL 60510, USA,
\ensuremath{^{t}}Nagasaki Institute of Applied Science, Nagasaki 851-0193, Japan,
\ensuremath{^{u}}National Research Nuclear University, Moscow 115409, Russia,
\ensuremath{^{v}}Northwestern University, Evanston, IL 60208, USA,
\ensuremath{^{w}}University of Notre Dame, Notre Dame, IN 46556, USA,
\ensuremath{^{x}}Universidad de Oviedo, E-33007 Oviedo, Spain,
\ensuremath{^{y}}CNRS-IN2P3, Paris, F-75205 France,
\ensuremath{^{z}}Universidad Tecnica Federico Santa Maria, 110v Valparaiso, Chile,
\ensuremath{^{aa}}The University of Jordan, Amman 11942, Jordan,
\ensuremath{^{bb}}Universite catholique de Louvain, 1348 Louvain-La-Neuve, Belgium,
\ensuremath{^{cc}}University of Z\"{u}rich, 8006 Z\"{u}rich, Switzerland,
\ensuremath{^{dd}}Massachusetts General Hospital, Boston, MA 02114 USA,
\ensuremath{^{ee}}Harvard Medical School, Boston, MA 02114 USA,
\ensuremath{^{ff}}Hampton University, Hampton, VA 23668, USA,
\ensuremath{^{gg}}Los Alamos National Laboratory, Los Alamos, NM 87544, USA,
\ensuremath{^{hh}}Universit\`{a} degli Studi di Napoli Federico I, I-80138 Napoli, Italy
}
\noaffiliation

\date{\today}
\begin{abstract}
Differential cross sections for the production
of $Z$ bosons or off-shell photons $\gamma^*$ in association with jets are measured in
proton-antiproton collisions at center-of-mass energy $\sqrt{s}=1.96$~TeV
using the full data set collected with the Collider Detector at
Fermilab in Tevatron Run II, and corresponding to
9.6~fb$^{-1}$ of integrated luminosity. Results include first
measurements at CDF of differential cross sections in events with a
\Zg{} boson and three or more jets, the inclusive cross section
for production of $Z/\gamma^*$ and four or more jets, and cross
sections as functions of various angular observables in lower
jet-multiplicity final states. Measured cross sections are compared to
several theoretical predictions.
\vspace*{3.0cm}

\end{abstract}


\maketitle

\section{\label{sec:Intro}Introduction}
Studies of the production of jets in association with a \Zg{} boson,
henceforth referred to as \Zjets{} processes, are central topics in
hadron collider physics. Differential cross section measurements
provide stringent tests for perturbative quantum chromodynamics (QCD)
predictions~\cite{Gross:1973ju}. In addition, \Zjets{} production is a
background to many rare standard model (SM) processes, such as
Higgs-boson production, and searches for non-SM physics. Dedicated
measurements can help to improve the theoretical modeling of \Zjets{}
production.

Differential cross sections have been previously measured in
proton-antiproton collisions by the CDF~\cite{Aaltonen:2007ae} and 
D0~\cite{Abazov:2008ez, *Abazov:2009av, *Abazov:2009pp} collaborations
as functions of several variables, including the jet transverse
momentum, the jet rapidity, and various angular observables. These
measurements are in qualitative agreement with predictions from
perturbative QCD at the next-to-leading order (NLO) expansion in the
strong-interaction coupling, but are limited by the small number of
events with high multiplicity of jets. Recently, measurements
have also been published by the ATLAS~\cite{Aad:2013ysa, *Aad:2011qv}
and CMS~\cite{Chatrchyan:2011ne, *Chatrchyan:2013tna, *Khachatryan:2014zya} collaborations
in proton-proton collisions at the LHC, since the understanding of
these SM processes is essential in the search for non-SM physics at
the LHC.

In this article, measurements of differential cross sections for
\Zjets{} production are presented, using the full data sample of
proton-antiproton collisions collected with the CDF II detector in Run
II of the Tevatron Collider, which corresponds to 9.6~fb$^{-1}$ of integrated luminosity.
The results include differential cross sections as functions of
jet transverse momentum, \pt, and rapidity, $y$~\footnote{The
  rapidity is defined as $y=\frac{1}{2}\ln(\frac{E+p_Z}{E-p_Z})$;
  the transverse momentum and energy are defined by $\pt = p
  \sin{\theta}$ and $\et = E \sin{\theta}$}, extended for the first
time at CDF to the \Zthreejets{} final state; the total cross section
as a function of jet multiplicity up to four jets; and several
differential distributions for events with a \Zg{} boson and at least one or
two jets. Measurements are compared to NLO~\cite{Campbell:2002tg,
  Berger:2008sj} and approximate next-to-next-to-leading order (NNLO) perturbative QCD
predictions~\cite{Rubin:2010xp}, to NLO QCD predictions including NLO
electroweak corrections~\cite{Denner:2011vu}, 
and to distributions from various Monte Carlo (MC) generators that use
parton showers interfaced with fixed-order calculations~\cite{Mangano:2002ea, Alioli:2010qp}.

This paper is organized as follows: Section~\ref{sec:CDF} contains a
brief description of the CDF II detector. The data sample and the event
selection are presented in Sec.~\ref{sec:Data}. The MC
samples used across the analysis are listed in
Sec.~\ref{sec:MCsamples}. The estimation of the background
contributions is described in Sec.~\ref{sec:Backgrounds}. The
unfolding procedure is explained in Sec.~\ref{sec:unf}. The
systematic uncertainties are addressed in Sec.~\ref{sec:sys}. The
theoretical predictions are described in Sec.~\ref{sec:pred}. The
measured differential cross sections are shown and discussed in Sec.~\ref{sec:results}.
Section~\ref{sec:conclusion} summarizes the results.


\section{\label{sec:CDF} The CDF II Detector}
The CDF II detector, described in detail in Ref.~\cite{Abulencia:2005ix},
is composed of a tracking system embedded in a $1.4$~T
magnetic field, surrounded by electromagnetic and hadronic
calorimeters and muon spectrometers. The CDF experiment uses a
cylindrical coordinate system in which the $z$ axis lies along the
proton beam direction, $\phi$ is the azimuthal angle, and $\theta$ is
the polar angle, which is often expressed as pseudorapidity $\eta =
-\ln [\tan(\theta/2)]$. The tracking system includes a silicon
microstrip detector~\cite{Sill:2000zz} covering a pseudorapidity
range of $|\eta|<2$, which provides precise three-dimensional
reconstruction of charged-particle trajectories (tracks).
The silicon detector is surrounded by a $3.1$~m long
open-cell drift chamber~\cite{Affolder:2003ep}, 
which covers a pseudorapidity range $|\eta|<1$, providing efficient
pattern recognition and accurate measurement of the momentum of
charged particles. The calorimeter system is
arranged in a projective-tower geometry and measures energies
of photons and electrons in the $|\eta|<3.6$ range. The
electromagnetic calorimeter~\cite{Balka:1987ty,  Hahn:1987tx} is a lead-scintillator sampling calorimeter, which also contains
proportional chambers at a depth corresponding approximately to the
maximum intensity of electron showers. The hadronic
calorimeter~\cite{Bertolucci:1987zn} is an iron-scintillator sampling
calorimeter. The muon detectors~\cite{Ascoli:1987av}, located outside
the calorimeters, consist of  drift chambers and scintillation
counters covering a pseudorapidity range of $|\eta|<1.0$. Finally, the
luminosity is computed from the rate of inelastic \ppbar{} collisions
determined by the Cherenkov counters~\cite{Elias:1999qg} located close
to the beam pipe.

\section{\label{sec:Data}Data Sample and Event Selection}
The data sample consists of \Zee{} and \Zmm{} + jets candidate events,
which have been collected using a three-level online event selection
system (trigger)~\cite{Winer:2001gj} between February 2002 and September
2011. In the electron channel, the trigger requires a central
(\mbox{$|\eta|\leqslant1$}) electromagnetic calorimeter cluster with \mbox{$\et{}\geqslant 18$}~GeV
matched to a charged particle with \mbox{$\pt\geqslant9$}~\gevc{}. 
In the analysis, \Zee{} events are selected by requiring two central
electrons with \mbox{$\et \geqslant 25$}~GeV and reconstructed invariant
mass in the range \mbox{$66 \leqslant M_{ee} \leqslant 116$}~\gevcsq{}. Details on the
electron identification requirements are given in
Ref.~\cite{Abulencia:2005ix}. In the muon channel, the trigger
requires a signal in the muon detectors associated with a charged
particle reconstructed in the drift chamber with \mbox{$|\eta|\leqslant1$} and
\mbox{$\pt \geqslant 18$}~\gevc{}. In the analysis, \Zmm{} events are selected
by requiring two reconstructed muons of opposite electric charge
with \mbox{$|\eta|\leqslant1$} and \mbox{$\pt \geqslant 25$}~\gevc{}, and reconstructed invariant
mass in the range \mbox{$66 \leqslant M_{\mu \mu} \leqslant 116$}~\gevcsq{}.
Quality requirements are applied to the tracks in order to reject misidentified
muons, and all the muon candidates are required to 
to be associated with an energy deposit in the calorimeter consistent
with a minimum ionizing particle. More details on the muon
reconstruction and identification can be found in
Ref.~\cite{Abulencia:2005ix}.

In addition to a \Z{} boson candidate, one or more jets with
\mbox{$\pt \geqslant 30$}~\gevc{} and rapidity \mbox{$|y|\leqslant2.1$} are required.
Jets are reconstructed using the midpoint
algorithm~\cite{Abulencia:2005yg} in a cone of radius \mbox{$R=0.7$}~\footnote{The jet cone radius $R$ is defined as $R =
  \sqrt{\eta^{2}+\phi^{2}}$}.
Calorimeter towers are clustered
if the energy deposits correspond to a transverse energy larger
than 0.1~GeV~\footnote{The transverse energy is evaluated using the
  position of the tower with respect to the primary interaction vertex.} and
used as seeds if larger than 1~GeV. Towers associated with
reconstructed electrons and muons are excluded.
A split-merge procedure is used, which merges a pair of cones if the
fraction of the softer cone's transverse momentum shared with the
harder cone is above a given threshold; otherwise the shared
calorimeter towers are assigned to the cone to which they are
closer. The split-merge threshold is set to 0.75.
Jet four-momenta are evaluated by adding the four-momenta of the
towers according to the E-scheme, $p^{\mu}_{\textrm{jet}} = \sum{p^{\mu}_{\textrm{towers}}}$, described in Ref.~\cite{Blazey:2000qt}.
With such a recombination scheme, jets are in general massive, and in
order to study the jet kinematic properties, the variables \pt{}
and $y$ are used, which account for the difference between $E$ and $p$
due to the jet mass.
Since the jet transverse momentum measured by the calorimeter,
\ptcal, is affected by instrumental effects, an average
correction~\cite{Bhatti:2005ai} is applied to \ptcal. These effects,
mainly due to the noncompensating nature of the calorimeter and the presence of
inactive material, are of the order of 30\% for \ptcal{} around
40~\gevc{} and reduce to about 11\% for high \ptcal{} jets. A
further correction is applied to account for the energy contributions
to jets from multiple \ppbar{} interactions, but no modification is made
to account for underlying-event contributions or fragmentation
effects. The requirement of \mbox{$\pt \geqslant 30$}~\gevc{} is applied to the
corrected jet transverse momentum. Events are selected if the leptons
are separated from the selected jets by $\Delta{R}_{\ell-\textrm{jet}} \geqslant
0.7$~\footnote{$\Delta{R}$ is defined as $\Delta{R} =
  \sqrt{\Delta{y}^{2}+\Delta{\phi}^{2}}$}.

\section{\label{sec:MCsamples}Monte Carlo Simulation}
Samples of \Zeejets{}, \Zmmjets{}, and \Zttjets{} events are generated using \alpgen{}
v2.14~\cite{Mangano:2002ea} interfaced to \pythia{}
6.4.25~\cite{Sjostrand:2006za} for the parton shower, with CTEQ5L
parton distribution functions (PDF)~\cite{Lai:1999wy} and using the
set of \emph{tuning} parameters denoted as Tune Perugia 2011~\cite{Skands:2010ak}.
The MLM matching procedure~\cite{Alwall:2007fs} is applied to avoid
double-counting of processes between the matrix-element calculations and the
parton-shower algorithm of \pythia. In addition, samples of
\ttbar, associated production of \W{} and \Z{}
bosons ($WW$, $WZ$, $ZZ$), and inclusive \Zg{} production are generated
using \pythia{} v6.2 with the same PDF set and Tune
A~\cite{Affolder:2001xt}. All the samples are passed through a full
CDF II detector simulation based on \textsc{geant}~\cite{Brun:1987ma},
where the \textsc{gflash}~\cite{Grindhammer:1989zg} package is used for
parametrization of the energy deposition in the calorimeters, and
corrected to account for differences between data and simulation in
the trigger selection and lepton identification efficiencies.
The electron \et{} and the muon \pt{} scale and resolution
are corrected to match the dilepton invariant mass distributions $M_{\ell\ell}$
observed in the data in the region \mbox{$84 \leqslant M_{\ell\ell} \leqslant 98$}~\gevcsq{}.
Simulated \Zjets{} samples are also reweighted with respect to the number of multiple
\ppbar{} interactions in the same bunch crossing so as to have the
same instantaneous luminosity profile of the data.
The MC samples are used to determine background
contributions and derive the unfolding correction factors described
in Sec.~\ref{sec:unf}.

\section{\label{sec:Backgrounds}Background Contributions}
The selected sample of \Zjets{} data events is expected to include
events from various background processes. The largest background
contributions come from pair production of \W{} and \Z{}
bosons, $WW$, $WZ$, $ZZ$, and top-antitop quarks, \ttbar; a
smaller contribution comes from \Zttjets{} events. Inclusive jets
and \Wjets{} events contribute to the background if one or more
jets are misidentified as electrons or muons. Various strategies are
used to estimate the background contributions. In the \Zee{} channel,
a data-driven method is used to estimate the inclusive jets and
\Wjets{} background contribution. First, the probability for a jet to
pass the electron selection requirements is evaluated using an
inclusive jet data sample. This is denoted as \emph{fake} rate
and is parametrized as a function of the jet transverse
energy. The fake rate is applied to jets from a sample of
events with one reconstructed electron: for each event, all the
possible electron-jet combinations are considered as \Zg{} candidates,
the jet transverse energy is corrected to match on average the
corresponding electron energy, and all the electron-jet pairs that
fulfill the selection requirements are weighted with the corresponding
fake rate associated with the jet, and used to estimate the
background rate for each observed distribution.

In the muon channel, the \Wjets{} and inclusive jets processes constitute a source of background
if a track inside a jet is identified as a muon. To estimate this
background contribution, events containing muon pairs are
reconstructed following the analysis selection but requiring the
charge of the two muons to have the same electric charge.

The other background contributions, originating from \ttbar, associated production of \W{} and \Z{}
bosons ($WW$, $WZ$, $ZZ$), and \Ztt{}~+~jets, are estimated with simulated samples.
The \ttbar{} sample is normalized according to the approximate NNLO cross
section~\cite{oai:arXiv.org:0807.2794}, the $WW$, $WZ$ and $ZZ$
samples are normalized according to the NLO cross
sections~\cite{oai:arXiv.org:hep-ph/9905386}, and the \Ztt{} + jets
sample is normalized according to the \Z{} inclusive NNLO cross
section~\cite{Abulencia:2005ix}.
The total background varies from about $2\%$ to $6\%$ depending on jet multiplicity as shown
in Table~\ref{tab:backg}, which reports the sample composition per
jet-multiplicity bin in the electron and muon channels.
\begin{table}
    \caption{Estimated background contributions, background systematic uncertainties, and data yield for (a)
      \ZeeNjets{} and (b) \ZmmNjets{} channels, with the number of jets $N_{\textrm{jets}} \geqslant 1,2,3$, and $4$.}
  \begin{center}
    \label{tab:backg}
    \begin{tabular}{lcccc}
      \toprule
      \multicolumn{1}{c}{$\Zee{}$ + jets } & \multicolumn{4}{c}{Estimated events}  \\
      \colrule
      Backgrounds &  $\geqslant 1$ jet & $\geqslant 2$ jets  & $\geqslant 3$ jets &  $\geqslant 4$ jets \\
      \colrule
      QCD, $W$ + jets               & $  25.9 \pm 3.9$  & $ 4.0  \pm 0.6$  & $0.6 \pm 0.1$ & $\leqslant 0.1$\\
      $WW$, $ZZ$, $ZW$              & $  119  \pm 36$   & $ 43   \pm 13$   & $4.2 \pm 1.3$ & $ 0.3 \pm 0.1$\\
      \ttbar{}                      & $  45   \pm 13$   & $ 25.4 \pm 7.6$  & $2.9 \pm 0.9$ & $ 0.2 \pm 0.1$\\
      \Ztt{} + jets                 & $  7.2  \pm 2.2$  & $ 0.5  \pm 0.1$  & $<0.1$        & $<0.1$        \\
      \colrule
      Total background              & $ 197 \pm  38 $ & $ 73 \pm 15 $ & $ 7.8  \pm  1.5$ & $ 0.7 \pm 0.1$\\
      \colrule
      Data                          & $ 12910 $ & $ 1451 $ & $ 137 $ & $ 13 $\\
      \botrule
    \end{tabular}\\
    \subfigure[]{}
  \end{center}
  \begin{center}
    \begin{tabular}{lcccc}
      \toprule
      \multicolumn{1}{c}{$\Zmm{}$ + jets } & \multicolumn{4}{c}{Estimated events}  \\
      \colrule
      Backgrounds &  $\geqslant 1$ jet & $\geqslant 2$ jets  & $\geqslant 3$ jets &  $\geqslant 4$ jets \\
      \colrule
      QCD, $W$ + jets              & $  51   \pm 51$   & $ 18   \pm 18$   & $3   \pm 3$   & $ 1   \pm 1$  \\
      $WW$, $ZZ$, $ZW$             & $  190  \pm 57$   & $ 69   \pm 21$   & $6.7 \pm 2.0$ & $ 0.5 \pm 0.2$\\
      \ttbar{}                     & $  68   \pm 21$   & $ 38   \pm 12$   & $4.5 \pm 1.3$ & $ 0.5 \pm 0.1$\\
      \Ztt{} + jets                & $  9.4  \pm 2.8$  & $ 1.2  \pm 0.3$  & $\leqslant 0.1$& $< 0.1$       \\
      \colrule
      Total background             & $ 318 \pm  79 $   & $ 126 \pm 30 $   & $ 14.3  \pm  3.8$ & $ 2.0 \pm 1.0$\\
      \colrule
      Data                         & $ 19578$ & $ 2247 $ & $ 196$ & $ 13$\\  
      \botrule
    \end{tabular}\\
    \subfigure[]{}
  \end{center}
\end{table}

\begin{figure*}
  \subfigure[]{\includegraphics[width=\figsize]{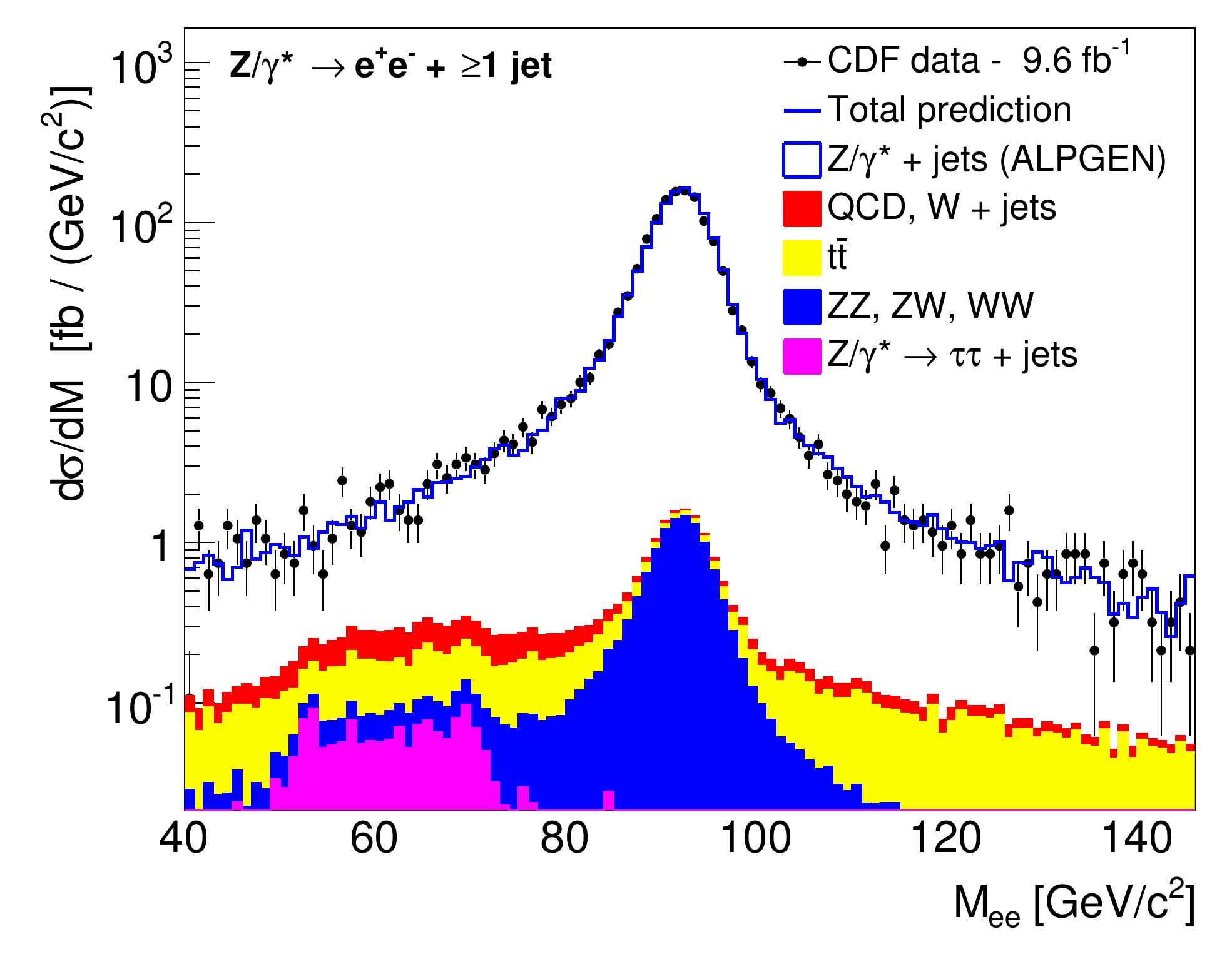}}
  \subfigure[]{\includegraphics[width=\figsize]{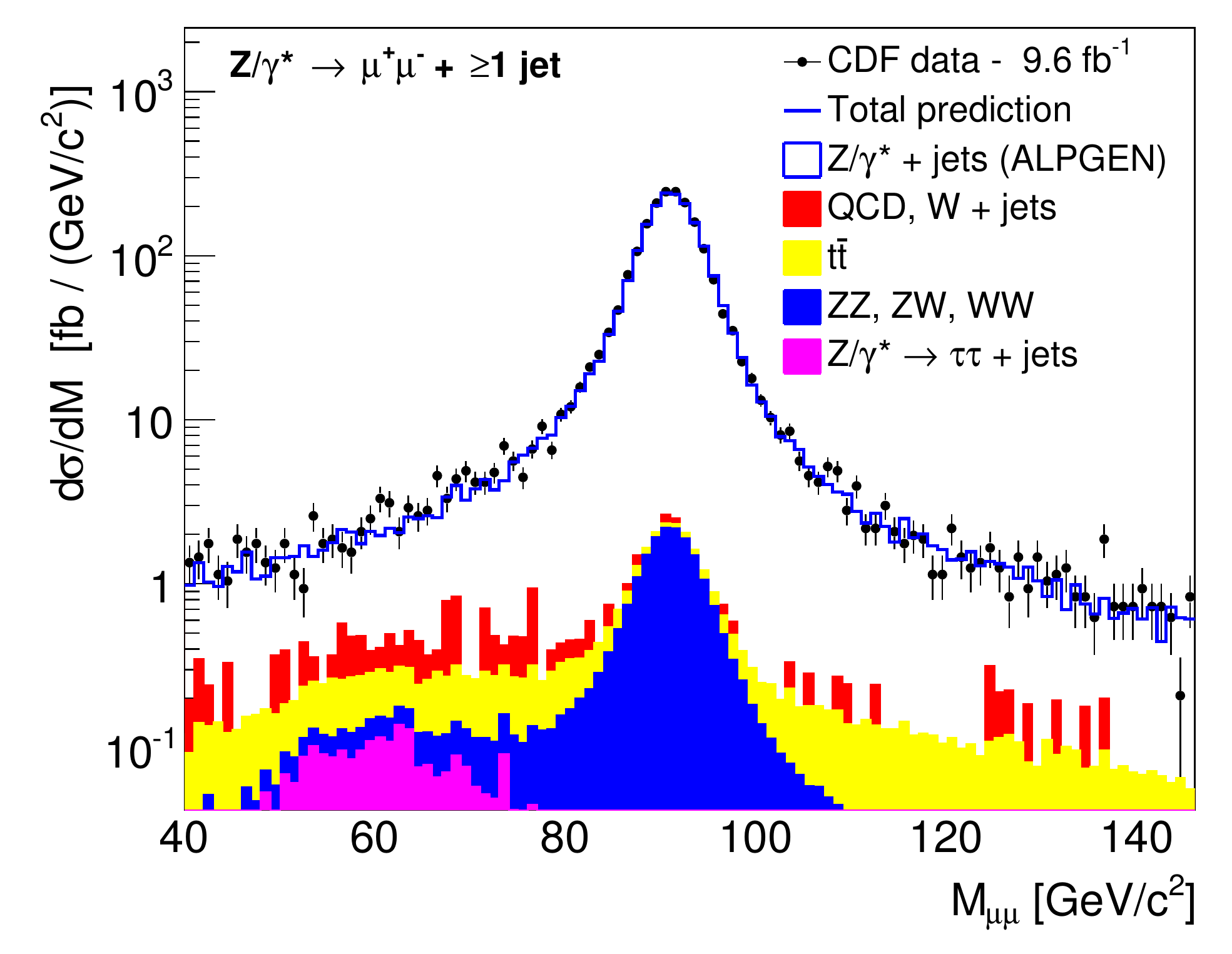}}
  \caption{Dilepton invariant mass distributions for events with at least one jet in
    the (a) \Zee{} and (b) \Zmm{} channels. Observed number of events divided by the integrated
    luminosity (black dots) are compared to the MC expectation
    (solid blue line), including signal and backgrounds contributions (filled
    histograms).
  }
  \label{fig:invmass}
\end{figure*}

Figure~\ref{fig:invmass} shows the invariant mass distribution for
\Zonejet{} events in the electron and muon decay channels. The
region outside the mass range used in the analysis contains a larger
fraction of background processes.
Table~\ref{tab:backg_sidebands} shows the comparison between data and
\begin{table*}
    \caption{Estimated background events and \Zjets{} MC prediction compared to the data in the 
      low- and high-mass regions outside the mass range used in the analysis,
      for $\Zee + \geqslant$ 1 jet and $\Zmm + \geqslant$ 1 jet events. Invariant mass ranges are given in \gevcsq{}.
      Background systematic uncertainties
      and statistical uncertainties of the \Zjets{} MC prediction are shown.}
  \begin{center}
    \label{tab:backg_sidebands}
    \begin{tabular}{lcccc}
      \toprule
                               & \multicolumn{2}{c}{$\Zee{} + \geq 1$ jet }                & \multicolumn{2}{c}{$\Zmm{} + \geq 1$ jet}  \\
      Backgrounds              &  $40 \leqslant M_{ee} < 66\quad$ &  $\quad116  < M_{ee} \leqslant 145\quad$ & $\quad40 \leqslant M_{\mu\mu} < 66\quad$ & $\quad116  < M_{\mu\mu} \leqslant 145$ \\
      \colrule                                                                                                                                     
      QCD, $W$ + jets          & $ 15.9  \pm  2.4$ &  $  2.9  \pm  0.4$ & $  37    \pm  37$   & $    8     \pm    8$   \\
      $WW$, $ZZ$, $ZW$         & $  5.2  \pm  1.6$ &  $  3.2  \pm  1.0$ & $   7.5  \pm   2.3$ & $    4.6   \pm    1.4$ \\
      \ttbar{}                 & $ 19.7  \pm  5.9$ &  $ 15.6  \pm  4.7$ & $  30.1  \pm   9.0$ & $   22.4   \pm    6.7$ \\
      \Ztt{} + jets            & $ 10.9  \pm  3.3$ &  $  0.3  \pm  0.1$ & $  17.5  \pm   5.2$ & $    0.3   \pm    0.1$ \\
      \colrule                                                                                                         
      Total background         & $ 51.7  \pm  7.3$ &  $ 21.9  \pm  4.8$ & $  92    \pm  39$   & $   35     \pm     11$ \\
      \Zg{} + jets (\alpgen{}) & $238.6  \pm  6.5$ &  $196.7  \pm  5.6$ & $ 335.4  \pm  7.2$  & $  289.0   \pm    6.4$ \\
      \colrule                                                                                                         
      Total prediction         & $290.3  \pm  9.8$ &  $218.6  \pm  7.3$ & $ 428    \pm  39$   &  $  324    \pm   12$  \\
      Data                     & $312$             &  $226$             & $ 486$              &  $  334$              \\
      \botrule
    \end{tabular}
  \end{center}
\end{table*}
\Zjets{} signal plus background prediction for \Zonejet{} events in
the low- and high-mass regions $40 \leqslant M_{\ell\ell} <
66$~\gevcsq{} and $116 < M_{\ell\ell} \leqslant 145$~\gevcsq{},
respectively. The good agreement between data and expectation supports
the method used to estimate the sample composition.

\section{\label{sec:unf}Unfolding}
Measured cross sections need to be corrected for detector effects in
order to be compared to the theoretical predictions.
The comparison between data and predictions is performed at the particle
level, which refers to experimental signatures reconstructed from
quasi-stable (lifetime greater than 10 ps) and color-confined final-state
particles including hadronization and underlying-event contributions, but not the
contribution of multiple \ppbar{} interactions in the same bunch
crossing~\cite{Buttar:2008jx}. Detector-level cross sections are calculated by subtracting the
estimated background from the observed events and dividing
by the integrated luminosity. Measured cross sections are unfolded
from detector level to particle level with a bin-by-bin procedure.
For each bin of a measured observable $\alpha$, the \alpgenpythia{}
\Zee{} + jets and \Zmm{} + jets MC samples are used to
evaluate the unfolding factors, which are defined as
$U_{\alpha}=\frac{d\sigma^{\textrm{MC}}_{\textrm{p}}}{d\alpha}/\frac{d\sigma^{\textrm{MC}}_{\textrm{d}}}{d\alpha}$, where
$\sigma^{\textrm{MC}}_{\textrm{p}}$ and $\sigma^{\textrm{MC}}_{\textrm{d}}$ are the
simulated particle-level and detector-level cross sections, respectively.
Measured particle level cross sections are evaluated as
$\frac{d\sigma_{\textrm{p}}}{d\alpha} = \frac{d\sigma_{\textrm{d}}}{d\alpha}
\cdot U_{\alpha}$, where $\sigma_{\textrm{d}}$ is the detector-level
measured cross section.
The simulated samples used for the unfolding are validated by
comparing measured and predicted cross 
sections at detector level. The unfolding factors account
for \Zll{} reconstruction efficiency, particle detection, and jet
reconstruction in the calorimeter. Unfolding factors are typically
around 2.5 (1.7) in value and vary between 2.3 (1.6) at low \pt{} and
3 (2) at high \pt{} for the \Zee{} (\Zmm) channel.

At particle level, radiated photons are recombined with leptons
following a scheme similar to that used in Ref.~\cite{Denner:2011vu}. A
photon and a lepton from \Zll{} decays are recombined when
$\Delta{R}_{\gamma-\ell} \leqslant 0.1$. If both charged leptons in the final
state are close to a photon, the photon is recombined with the lepton with the
smallest $\Delta{R}_{\gamma-\ell}$. Photons that are not recombined to
leptons are included in the list of particles for the jet
clustering. With such a definition, photons are clustered
into jets at the particle level, and \Zg{} + $\gamma$ production is
included in the definition of \Zjets{}. The contribution of the \Zg{} +
$\gamma$ process to the \Zjets{} cross section is at the percent
level, and taken into account in the \pythia{} simulation through photon
initial- (ISR) and final-state radiation (FSR).

Reconstruction of experimental signatures and kinematic requirements
applied at particle level establish the measurement definition.
Requirements applied at the detector level are also applied to jets
and leptons at the particle level so as to reduce the uncertainty of
the extrapolation of the measured cross section. Jets are
reconstructed at particle level in the simulated sample with the
midpoint algorithm in a cone of radius $R=0.7$, the split-merge
threshold set to 0.75, and using as seeds particles with $\pt
\geqslant 1$ \gevc{}. The measured cross sections are defined in the
kinematic region \mbox{$66 \leqslant M_{\ell\ell} \leqslant
  116$}~\gevcsq{}, \mbox{$|\eta^{\ell}|\leqslant1$}, 
\mbox{$\pt^{\ell}  \geqslant 25$}~\gevc{} ($\ell=e,~\mu$), 
\mbox{$\ptjet \geqslant 30$}~\gevc{}, \mbox{$|\yjet|\leqslant2.1$}, and
\mbox{$\Delta{R}_{\ell-\textrm{jet}} \geqslant 0.7$}.

\section{\label{sec:sys}Systematic Uncertainties}
All significant sources of systematic uncertainties are studied. The
main systematic uncertainty of the \Zll{} + jets measurement is
due to the jet-energy-scale correction. The jet-energy scale is
varied according to Ref.~\cite{Bhatti:2005ai}. Three sources of
systematic uncertainty are considered: the absolute jet-energy scale, multiple \ppbar{} interactions,
and the $\eta$-dependent calorimeter response. The absolute jet-energy
scale uncertainty depends on the response of the calorimeter to
individual particles and on the accuracy of the simulated model for
the particle multiplicity and \pt{} spectrum inside a jet. 
This uncertainty significantly affects observables involving high-\pt{} jets
and high jet multiplicity. The jet-energy uncertainty
related to multiple \ppbar{} interactions arises from inefficiency in
the reconstruction of multiple interaction vertices, 
and mainly affects jets with low \pt{} and high rapidity, and
events with high jet multiplicity. The $\eta$-dependent uncertainty
accounts for residual discrepancies between data and simulation after
the calorimeter response is corrected for the dependence on $\eta$.

Trigger efficiency and lepton identification uncertainties are of the
order of $1\%$ and give small contributions to the total uncertainty.

\begin{figure*}
    \centering
    \subfigure[]{\includegraphics[width=\figsize]{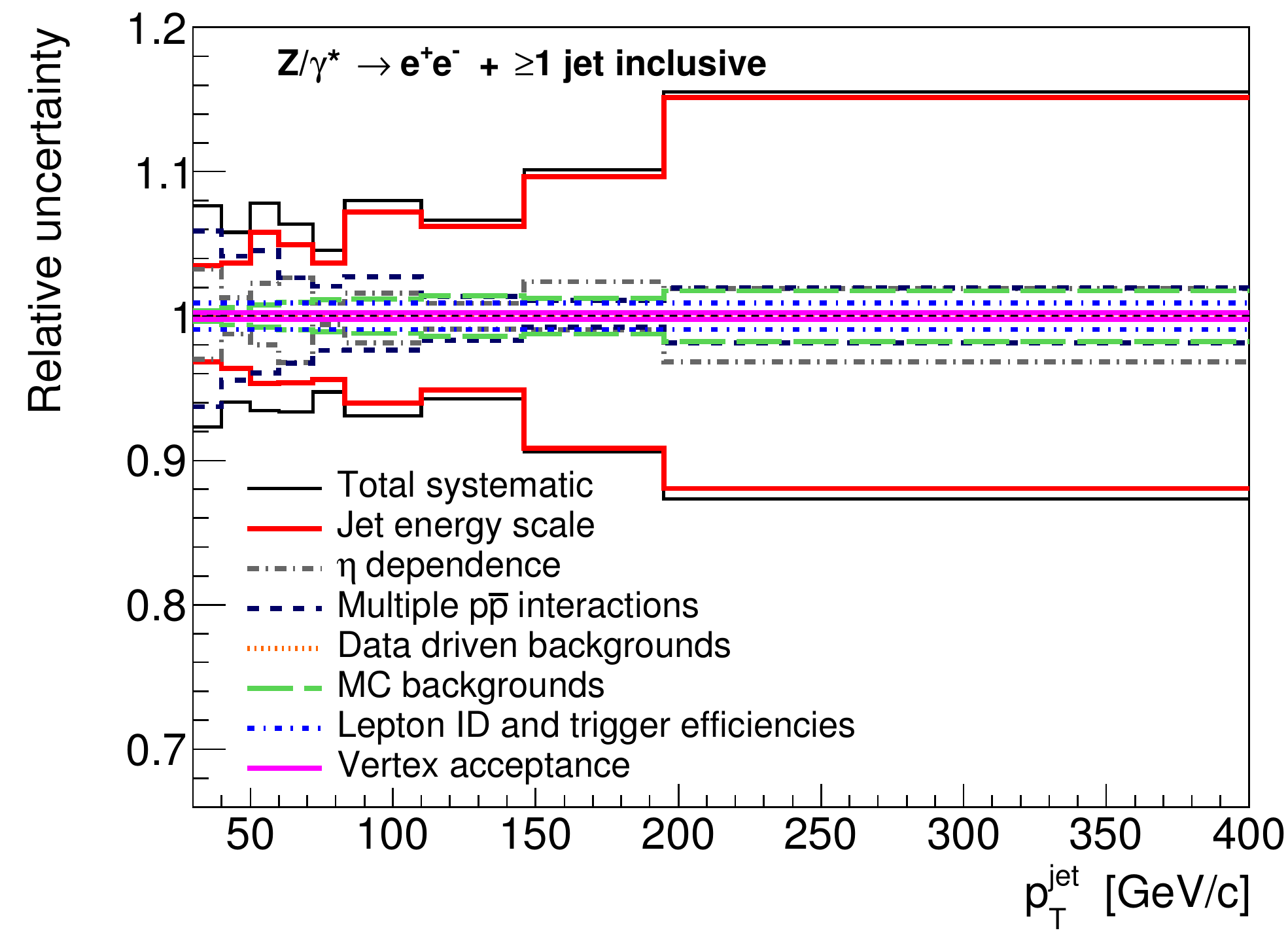}}
    \subfigure[]{\includegraphics[width=\figsize]{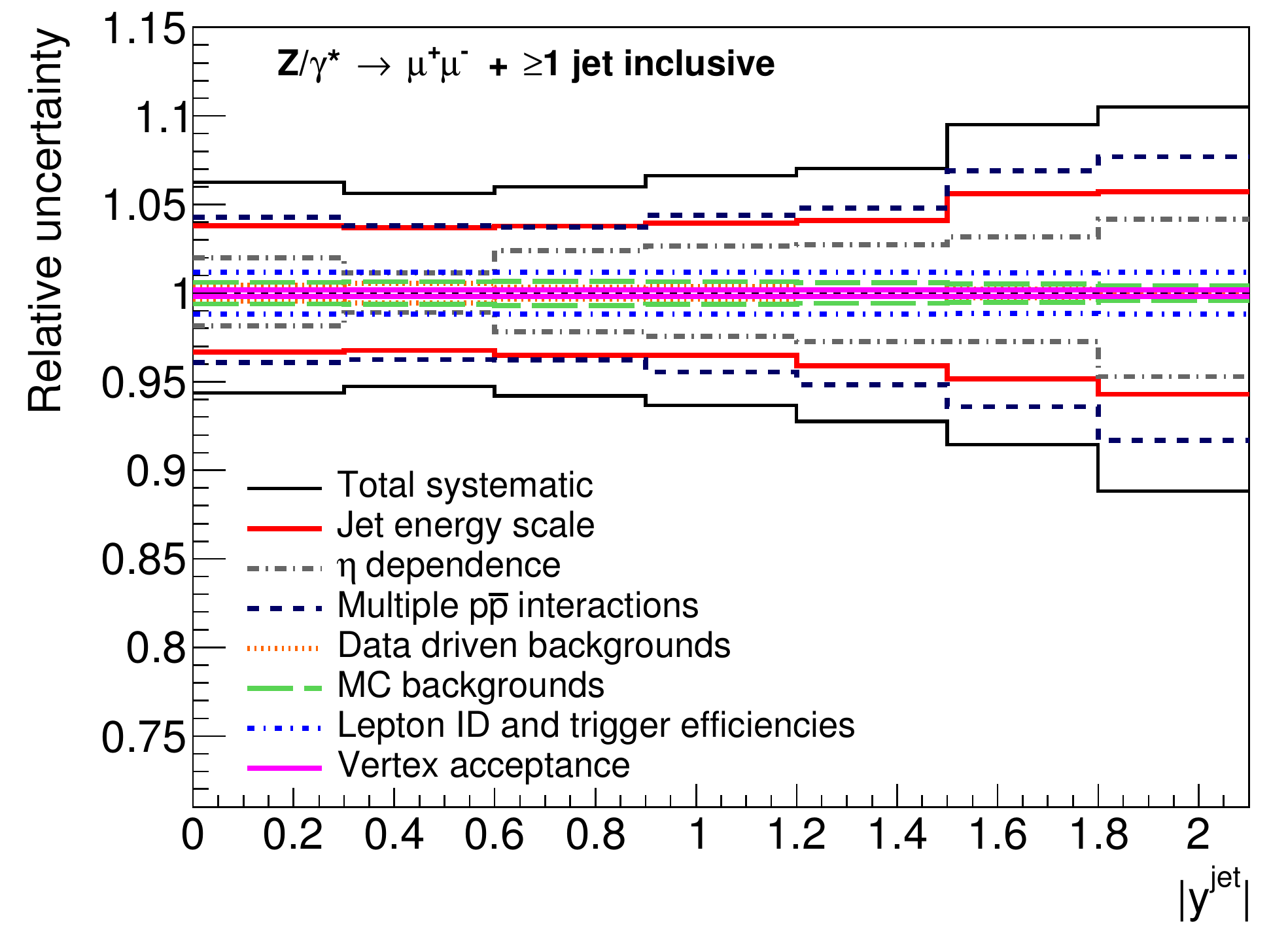}}
    \caption{Relative systematic uncertainties as functions of (a)
      inclusive jet \pt{} in the \Zee{} channel and (b) inclusive jet
      rapidity in the \Zmm{} channel, for events with \Zonejet.
      \label{fig:Total_Sys_1J}}
\end{figure*}

A $30\%$ uncertainty is applied to the MC
backgrounds yield estimation, to account for missing higher-order
corrections on the cross-section normalizations~\cite{Aaltonen:2007ae}. In the \Zee{}
channel, a $15\%$ uncertainty is assigned to the data-driven QCD and
\Wjets{} background yield estimation, to account for the statistical and
systematic uncertainty of the fake-rate parametrization. In the \Zmm{}
channel a $100\%$ uncertainty is applied to the subtraction of QCD and
\Wjets{} background, which accounts for any difference between the
observed same-charge yield and the expected opposite-charge background
contribution. The impact of both sources to the uncertainties of the
measured cross sections is less than $2\%$. The primary vertex
acceptance is estimated by fitting the beam luminosity as a function of
$z$ using minimum bias data, the uncertainty on the primary vertex
acceptance is approximately $1\%$. Finally, the luminosity
estimation has an uncertainty of $5.8\%$ which is applied to the
measurements~\cite{Klimenko:2003if}. As examples, systematic uncertainties as
functions of inclusive jet \pt{} in the \Zee{} channel and inclusive
jet rapidity in the \Zmm{} channel are shown in
Fig.~\ref{fig:Total_Sys_1J},
the corresponding systematic uncertainties as functions of inclusive
jet \pt{} in the \Zmm{} channel and inclusive jet rapidity in the
\Zee{} channel have similar trends.

\section{\label{sec:pred}Theoretical Predictions}
Measured \Zjets{} differential cross sections are compared to several
theoretical predictions such as NLO perturbative QCD calculations
evaluated with \mcfm~\cite{Campbell:2002tg} and
\blackhatsherpa~\cite{Berger:2008sj}, approximate NNLO
\loopsimmcfm{} predictions~\cite{Rubin:2010xp}, perturbative NLO QCD predictions
including NLO electroweak corrections~\cite{Denner:2011vu}, and to generators based on LO matrix
element (ME) supplemented by parton showers (PS), like
\alpgenpythia~\cite{Mangano:2002ea, Sjostrand:2006za}, and NLO
generators interfaced to PS as \powhegpythia~\cite{Alioli:2010qp}. For the \loopsimmcfm{}
predictions, the notation $\overline{\textrm{n}}^p$N$^q$LO introduced in Ref.~\cite{Rubin:2010xp} is used, 
which denotes an approximation to the N$^{p+q}$LO result in which the
$q$ lowest loop contributions are evaluated exactly, whereas the
$p$ highest loop contributions are evaluated with the \loopsim{}
approximation; according to such a notation, the approximate NNLO
\loopsimmcfm{} predictions are denoted with \nnlo{}.
The NLO \mcfm{} predictions are available for final states from \Zg{}
production in association with one or more, and two or more jets,
\loopsimmcfm{} only for the \Zonejet{} final state, NLO
\blackhatsherpa{} for jet multiplicity up to \Zthreejets{}, and
\powhegpythia{} predictions are available for all jet multiplicities
but have NLO accuracy only for \Zonejet. The \alpgen{} LO calculation
is available for jet multiplicities up to \Zg{} + 6 jets but, for the
current comparison, the calculation is restricted to up to
\Zfourjets. Electroweak corrections at NLO are available for the
\Zonejet{} final state.
Table~\ref{tab:theory_predictions} lists the theoretical predictions
which are compared to measured cross sections.
\begin{table*}
    \caption{Summary of the theoretical predictions compared to the measured cross sections. 
The order of the expansion in the strong-interaction coupling (QCD order), the order of the expansion in the fine-structure constant (EW order), the matching
to a parton shower, and the available jet multiplicities in \Zjets{} production are shown for each prediction.
 }
  \begin{center}
    \label{tab:theory_predictions}
    \begin{tabular}{lllll}
      \toprule
      Prediction               & QCD order  & EW order          & Parton shower & Jets multiplicity \\
      \colrule
      \mcfm{}                  & LO/NLO     & LO                & no            & $\Zg{} + \geqslant 1$ and $2$ jets \\
      \blackhatsherpa{}        & LO/NLO     & LO                & no            & $\Zg{} + \geqslant 1, 2$, and 3 jets \\
      \loopsimmcfm{}           & \nnlo{}    & LO                & no            & $\Zg{} + \geqslant 1$ jet \\
      \nloqcdew{} & NLO        & NLO               & no            & $\Zg{} + \geqslant 1$ jet \\
      \alpgenpythia{}          & LO         & LO                & yes           & $\Zg{} + \geqslant 1,2,3$, and $4$ jets \\
      \powhegpythia{}          & NLO        & LO                & yes           & $\Zg{} + \geqslant 1,2,3$, and $4$ jets \\
      \botrule
    \end{tabular}
  \end{center}
\end{table*}

The input parameters of the various predictions are chosen to be
homogeneous in order to emphasize the difference between
the theoretical models. The MSTW2008~\cite{Martin:2009iq} PDF sets
are used as the default choice in all the predictions.
The LO PDF set and one-loop order for the running of the strong-interaction coupling constant
$\alpha_s$ are used for the LO \mcfm{} and \blackhatsherpa{}
predictions; the NLO PDF set and two-loop order for the running of $\alpha_s$
for \powheg{}, \alpgen{}, NLO \mcfm{}, and NLO \blackhat{} predictions;
the NNLO PDF set and three-loop order for the running of $\alpha_s$
for the \nnlo{} \loopsim{} prediction. The contribution to the NLO
\mcfm{} prediction uncertainty due to the PDF is estimated with the
MSTW2008NLO PDF set at the 68\% confidence level (CL), by using the
Hessian method~\cite{Pumplin:2001ct}. There are 20
eigenvectors and a pair of uncertainty PDF associated with each
eigenvector. The pair of PDF corresponds to positive and negative 68\%
CL excursions along the eigenvector. The PDF contribution to the
prediction uncertainty is the quadrature sum of prediction
uncertainties from each uncertainty PDF.
The impact of different PDF sets is studied in \mcfm{},
\alpgen{} and \powheg{}. The variation in the predictions with
CTEQ6.6~\cite{Nadolsky:2008zw}, NNPDF2.1~\cite{Ball:2011mu},
CT10~\cite{Lai:2010vv}, and MRST2001~\cite{Martin:2001es} PDF sets is
of the same order of the MSTW2008NLO uncertainty. The LHAPDF 5.8.6
library~\cite{Whalley:2005nh} is used to access PDF sets, except
in \alpgen{}, where PDF sets are provided within the MC program.

The nominal choice~\cite{Berger:2010vm, *Berger:2009ep, Bauer:2009km}
for the functional form of the renormalization and factorization
scales is $\mu_0=\hat{H}_T/2=\frac{1}{2}\big(\sum_j p_{T}^{j} +
p_{T}^{\ell^+} + p_{T}^{\ell^-}\big)$~\footnote{In \blackhat{} and \powheg{}
  predictions, the alternative definition
  $\mu_0=\hat{H'}_{\textrm{T}}/2=\frac{1}{2}\big(\sum_j p_{\textrm{T}}^{j} + E_{\textrm{T}}^{Z}\big)$ 
  with $E_{\textrm{T}}^{Z} = \sqrt{M_{Z}^{2} + p_{\textrm{T},Z}^{2}}$ is
  used, where the index $j$ runs over the partons in the final state.},
where the index $j$ runs over the partons in the final state. An
exception to this default choice is the \alpgen{} prediction, which
uses $\mu_0=\sqrt{m_{Z}^{2} + \sum_j p_{T}^{j}}$; the difference with
respect to $\mu_0=\hat{H}_T/2$ was found to be
negligible~\cite{Camarda:2012yha}. The factorization and
renormalization scales are varied simultaneously between half and
twice the nominal value $\mu_0$, and the corresponding variations in
the cross sections are considered as an uncertainty of the
prediction. This is the largest uncertainty associated with the
theoretical models, except for the \alpgenpythia{} prediction, where
the largest uncertainty is associated with the variation of the
renormalization scale using the Catani, Krauss, Kuhn, Webber (CKKW) scale-setting
procedure~\cite{Catani:2001cc}. In the \alpgen{} prediction, the value of
the QCD scale, $\Lambda_{QCD}$, and the running order of the strong-interaction coupling constant
in the CKKW scale-setting procedure, $\alpha_{s}^{\textrm{CKKW}}$,
are set to $\Lambda_{QCD}=0.26$ and one loop, respectively~\cite{Cooper:2011gk}.
These settings match the corresponding values of
$\Lambda_{QCD}$ and the running order of $\alpha_s$ for
ISR and FSR of the \pythia{} Tune Perugia 2011. The variation of the
CKKW renormalization scale is introduced together with an opposite
variation of $\Lambda_{QCD}$ in the
\pythia{} tune. Simultaneous variations of the renormalization and factorization
scales for the matrix element generation in \alpgen{} were found to be smaller than
the variation of the CKKW scale~\cite{Camarda:2012yha}.
The differences with respect to the previously used Tune A and Tune
DW~\cite{Albrow:2006rt,*Field:2006gq} are studied, with the $\alpha_s$-matched setup of Tune Perugia 2011
providing a better modeling of the shape and normalization of the
\Zjets{} differential cross sections. In the case of Tune A and Tune DW,
the running of $\alpha_{s}^{\textrm{CKKW}}$ in \alpgen{} and $\Lambda_{QCD}$ in
\pythia{} is determined by the PDF set, which is CTEQ5L in
both to avoid mismatch. The \powheg{}
calculation is performed with the weighted events option, and the Born
suppression factor for the reweight is set to $10$~\gevc{}, following
Ref.~\cite{Alioli:2010qp}. Further studies on the impact of different
choices of the functional form of the renormalization and
factorization scales have been performed in Ref.~\cite{Camarda:2012yha}.

In the LO and NLO \mcfm{} predictions, jets are clustered with the
native \mcfm{} \emph{cone} algorithm with $R=0.7$. 
This is a seedless cone algorithm that follows the jet
clustering outlined in Ref.~\cite{Blazey:2000qt}. The
split-merge threshold is set to 0.75,
and the maximum $\Delta{R}$ separation $R_{sep}$ for two partons
to be clustered in the same jet~\cite{Ellis:1992qq}, is
set to \mbox{$R_{sep}=1.3R$}~\cite{Aaltonen:2007ae}. For the 
\loopsimmcfm{} prediction the minimum jet \pt{} for the generation is
set to $1$ \gevc{}, and the jet clustering is performed with the
fastjet~\cite{Cacciari:2011ma} interface to the
SISCone~\cite{Salam:2007xv} jet algorithm with cone radius $R=0.7$ and
a split-merge threshold of 0.75. The same parameters and setup for the
jet clustering are used in the \blackhatsherpa{} calculation, and the
predictions are provided by the \blackhat{} authors.

A recently developed MC program allows the calculation of
both NLO electroweak and NLO QCD corrections to the \Zonejet{} cross
sections~\cite{Denner:2011vu}. In such a prediction, the QCD and electroweak part of the NLO
corrections are combined with a factorization ansatz: NLO QCD
and electroweak corrections to the LO cross section are evaluated
independently and multiplied.
Such a combined prediction is referred to as \nloqcdew{}.
The prediction is evaluated with the configuration described
in Ref.~\cite{Denner:2011vu}, except for the renormalization and
factorization scales, which are set to $\mu_0=\hat{H}_T/2$, and the
predictions are provided by the authors.

\begin{figure*}
  \begin{center}
    \subfigure[]{\includegraphics[width=\figsize]{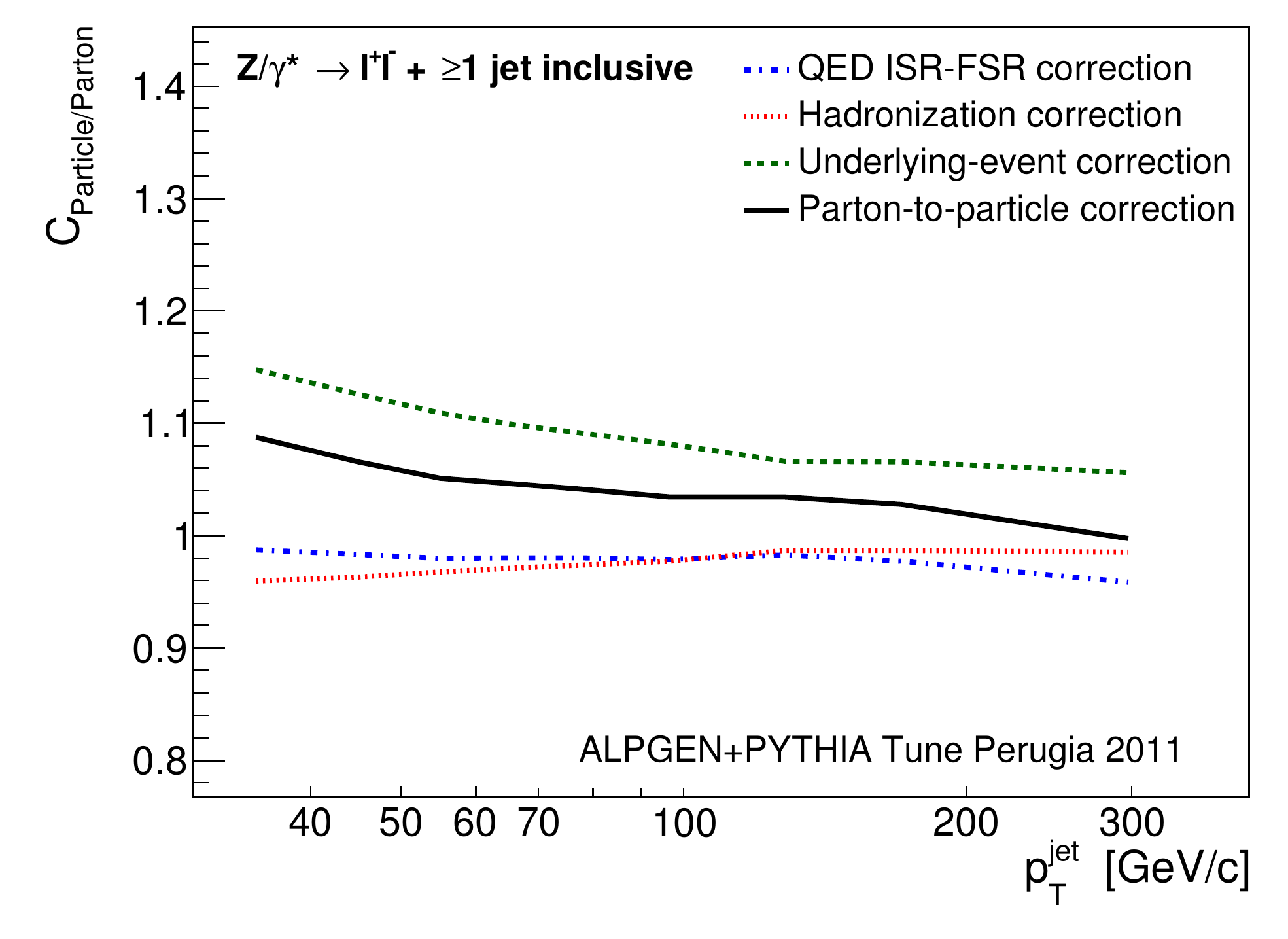}}
    \subfigure[]{\includegraphics[width=\figsize]{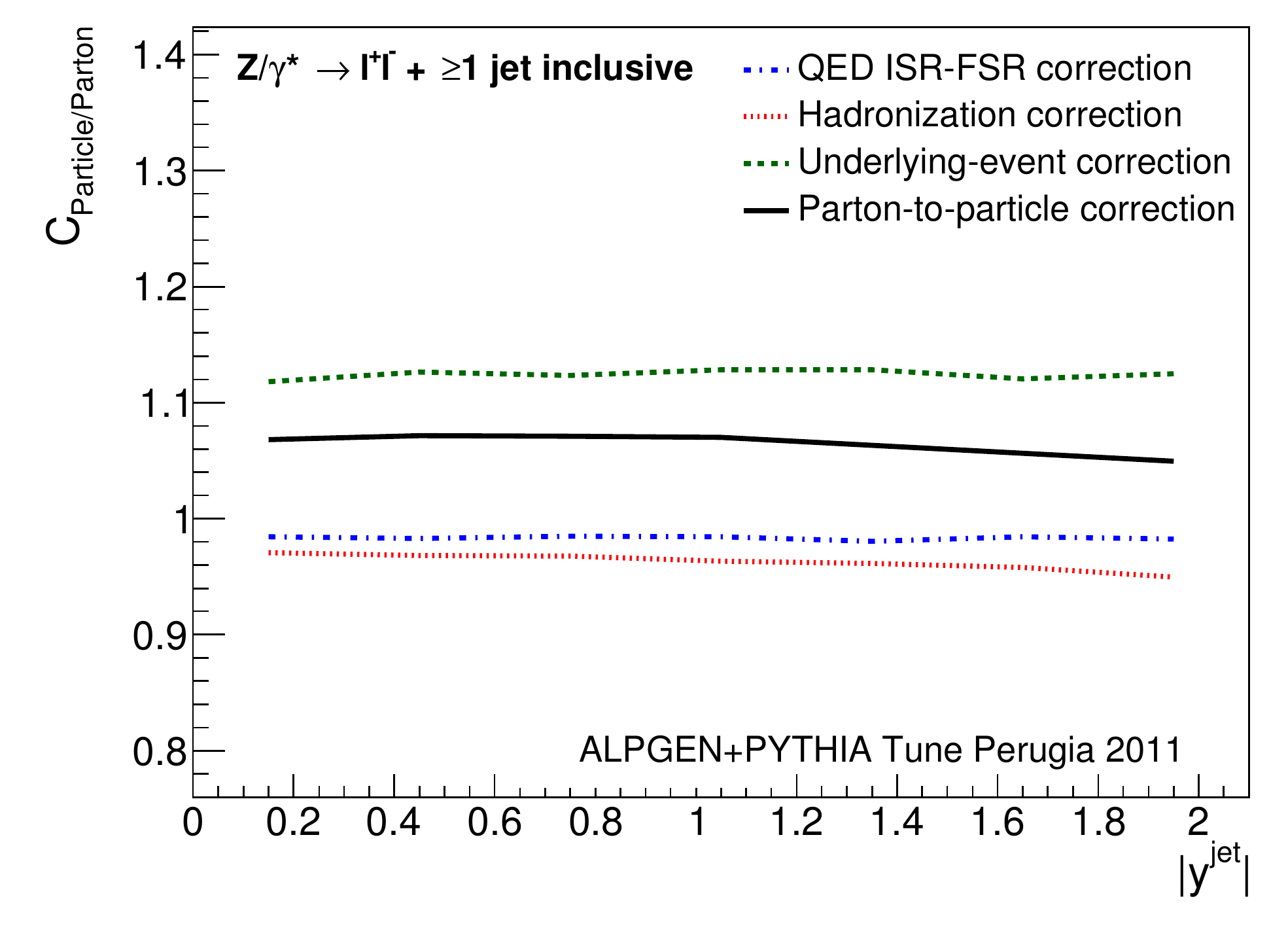}}
    \caption{Parton-to-particle corrections as functions of (a)
      inclusive jet \pt{} and (b) inclusive jet rapidity for \Zg{} + $\geqslant 1$ jet events.
The relative contributions of QED radiation, hadronization, and underlying event are shown.
      \label{fig:CC_Pt1Y1_P2H}}
  \end{center}
\end{figure*}

\begin{figure}
  \begin{center}
    \includegraphics[width=\figsize]{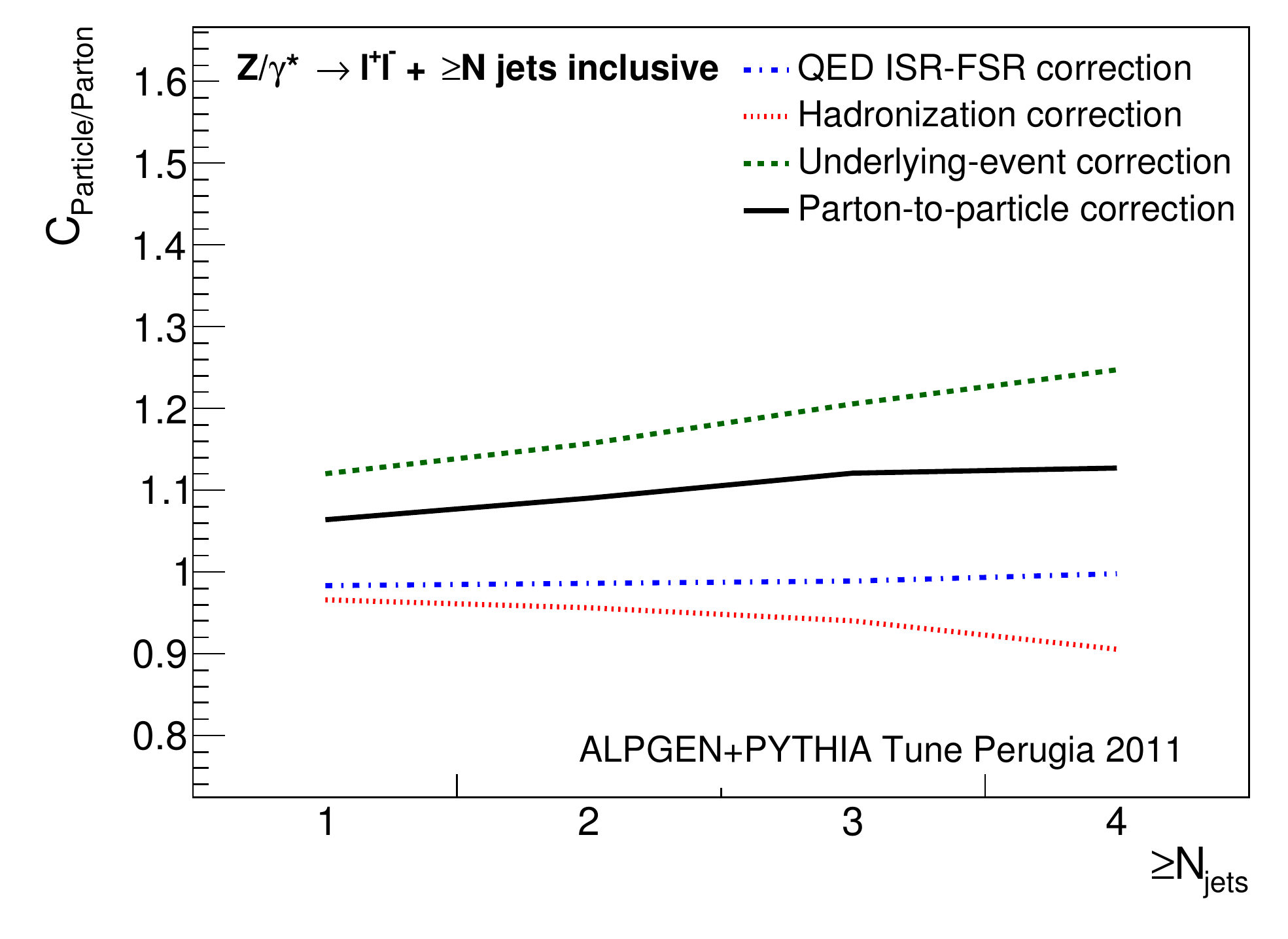}
    \caption{Parton-to-particle corrections as a function of jet
              multiplicity. The relative contributions of QED
              radiation, hadronization, and underlying event are shown.
      \label{fig:CC_NJ_Incl_P2H}}
  \end{center}
\end{figure}

\begin{figure*}
  \begin{center}
    \subfigure[]{\includegraphics[width=\figsize]{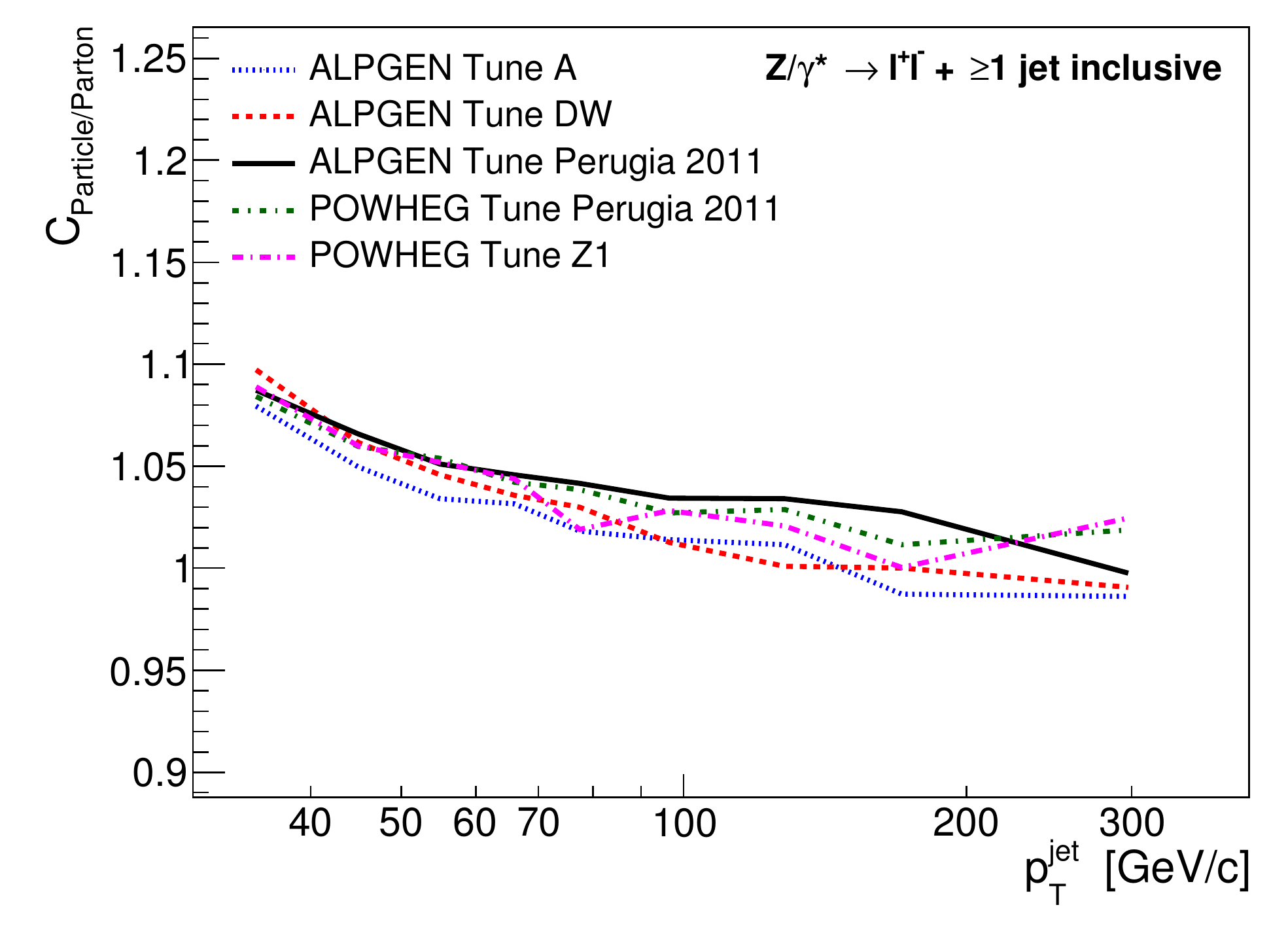}}
    \subfigure[]{\includegraphics[width=\figsize]{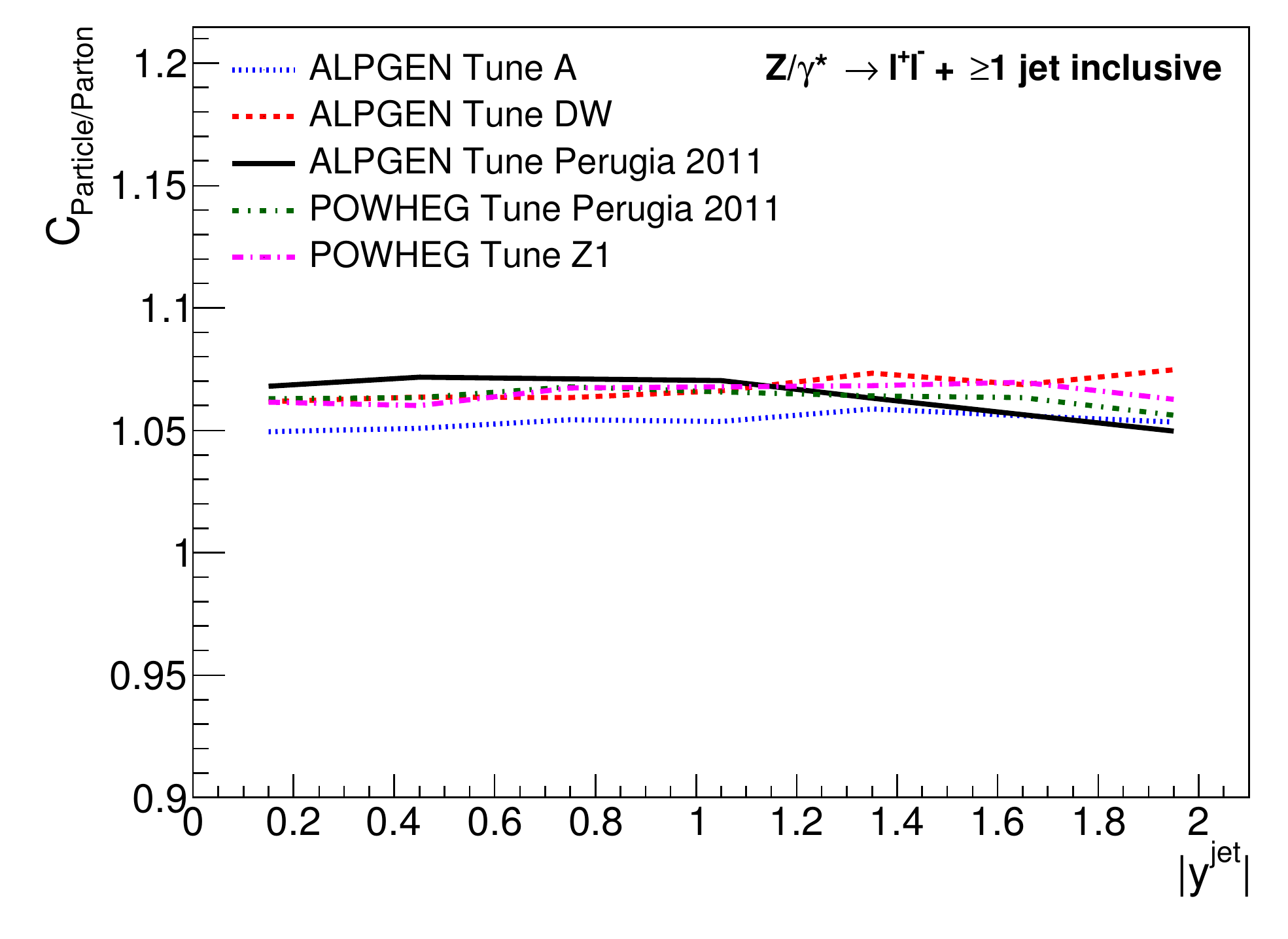}}
    \caption{Parton-to-particle corrections as functions of (a)
      inclusive jet \pt{} and (b) inclusive jet rapidity for \Zg{} + $\geqslant 1$ jet events,
      with various choices of the \pythia{} tune and different matrix
      element generators \alpgen{} or \powheg{}.
      \label{fig:CC_Pt1Y1_TUNE}}
  \end{center}
\end{figure*}

Fixed-order perturbative QCD predictions need to be corrected for
nonperturbative QCD effects in order to compare them with the
measured cross sections, including the underlying event associated with
multiparton interactions, beam remnants, and hadronization. Another
important effect that is not accounted for in the perturbative QCD
predictions and needs to be evaluated is the quantum electrodynamics
(QED) photon radiation from leptons and quarks. Both ISR and FSR are
considered, with the main effect coming from FSR. The inclusion of QED radiation
also corrects the \Zjets{} cross sections for the contribution of
\Zg{} + $\gamma$ production, which enters the definition of the
\Zjets{} particle level used in this measurement. The nonperturbative
QCD effects and the QED radiation are estimated with the MC
simulation based on the $\alpha_s$-matched Perugia 2011 configuration
of \alpgenpythia{}, where \pythia{} handles the simulation of these
effects. To evaluate the corrections, parton-level and particle-level
\alpgenpythia{} cross sections are defined: parton-level cross
sections are calculated with QED radiation, hadronization, and
multiparton interactions disabled in the \pythia{} simulation, whereas
these effects are simulated for the particle-level cross
sections. Kinematic requirements on leptons and jets and
jet-clustering parameters for the parton and particle levels are the
same as those used for the measured cross sections, and photons are
recombined to leptons in $\Delta{R}=0.1$ if radiated photons are
present in the final state. The corrections are obtained by evaluating the ratio of the
particle-level cross sections over the parton-level cross sections, bin-by-bin for
the various measured variables. Figure~\ref{fig:CC_Pt1Y1_P2H} shows the parton-to-particle
corrections as functions of inclusive jet \pt{} and inclusive jet
rapidity for \Zonejet{} events, with the contributions from
QED ISR and FSR radiation, hadronization, and underlying event. The
corrections have a moderate dependence on jet multiplicity, as
shown in Fig.~\ref{fig:CC_NJ_Incl_P2H}. Figure~\ref{fig:CC_Pt1Y1_TUNE} shows
the parton-to-particle corrections evaluated with various tunes of
the underlying-event and hadronization model in \pythia{}, namely Tune
A~\cite{Affolder:2001xt}, Tune DW~\cite{Albrow:2006rt,*Field:2006gq},
Tune Perugia 2011~\cite{Skands:2010ak}, and Tune Z1~\cite{Field:2011iq}, and with
the \alpgenpythia{} or \powhegpythia{} simulations. The corrections
are generally below $10\%$, and independent of the \pythia{} MC tune
and of the underlying matrix-element generator.

\begin{figure*}
  \begin{center}
    \subfigure[]{\includegraphics[width=\figsize]{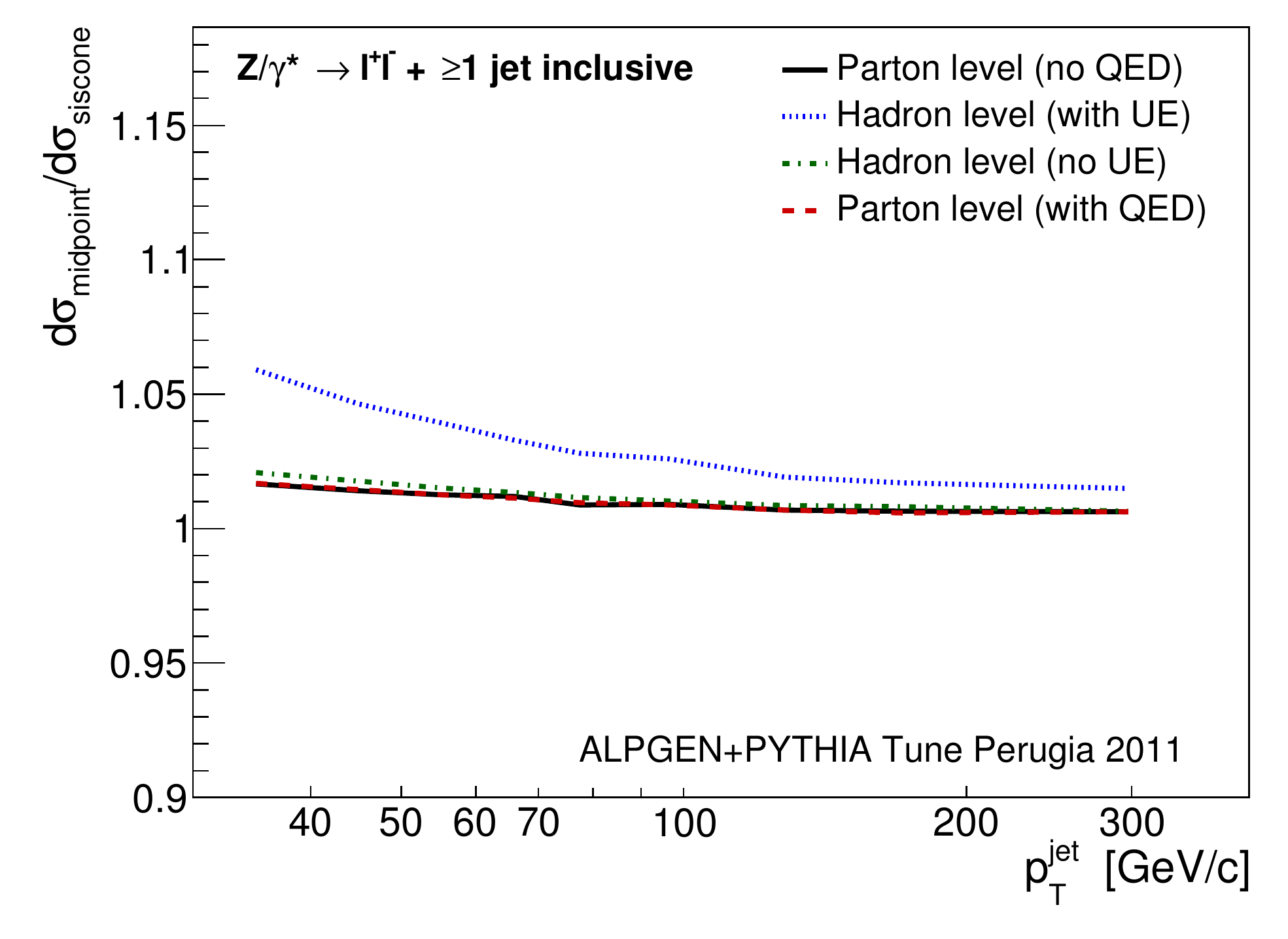}}
    \subfigure[]{\includegraphics[width=\figsize]{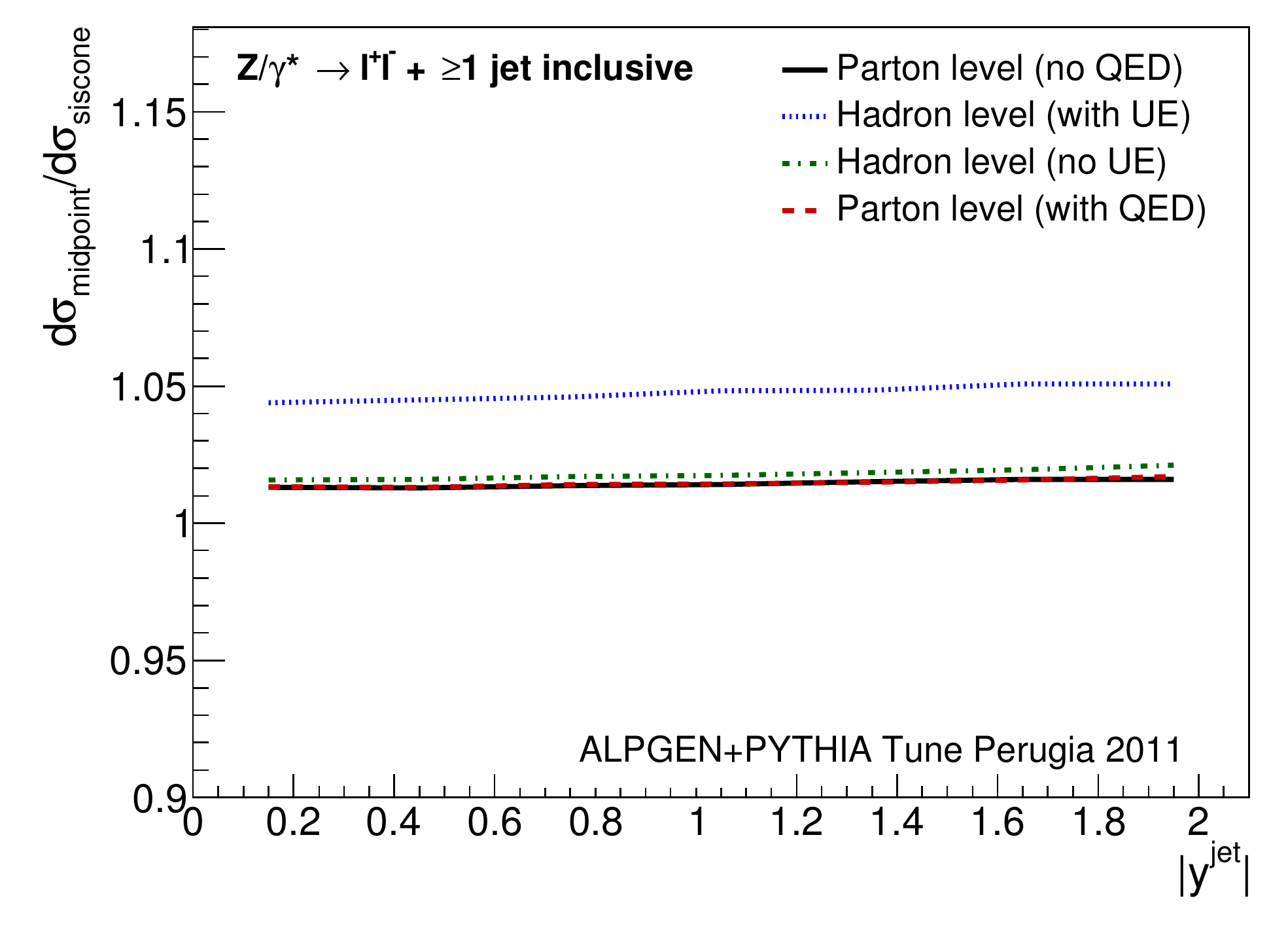}}
    \caption{Ratio of differential cross sections evaluated with the
      midpoint and with the SISCone jet algorithms, as functions
      of (a) inclusive jet \pt{} and (b) inclusive jet rapidity in
      \Zg{} + $\geqslant 1$ jet events.
      \label{fig:CC_Pt1_Incl_JETALG}}
  \end{center}
\end{figure*}

\begin{figure}
  \begin{center}
    \includegraphics[width=\figsize]{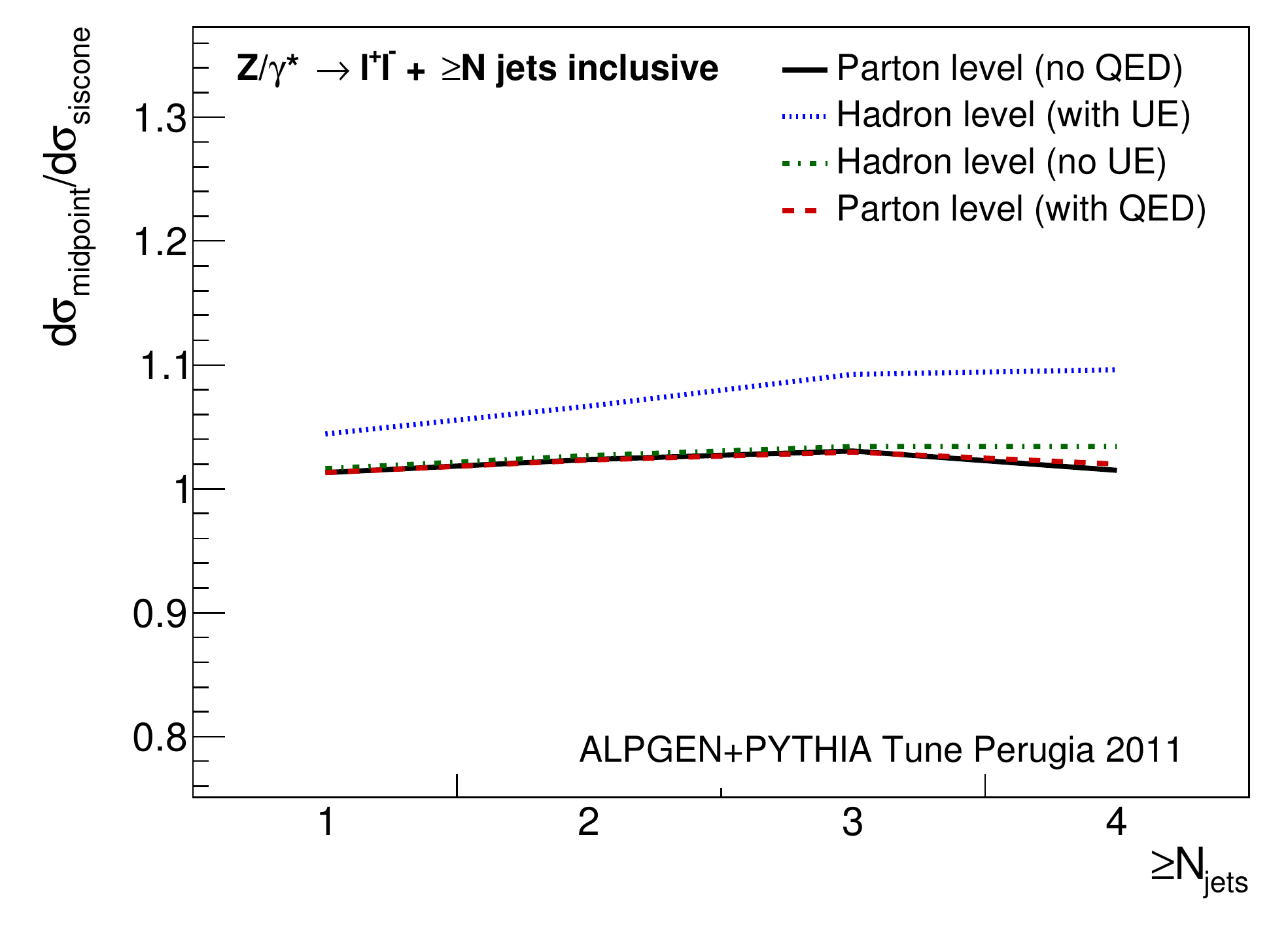}
    \caption{Ratio of differential cross sections evaluated with the
      midpoint and with the SISCone jet algorithms, as a function
      of jet multiplicity in \ZNjets{}.
      \label{fig:CC_NJ_Incl_JETALG}}
  \end{center}
\end{figure}

The \Zjets{} cross sections are measured using the midpoint
algorithm for the reconstruction of the jets in the final state.
The midpoint algorithm belongs to the class of iterative cone
algorithms. Though they present several experimental advantages, iterative cone
algorithms are not infrared and collinear safe,
which means that the number of hard jets found by such jet algorithms
is sensitive to a collinear splitting or to the addition of a soft emission.
In particular the midpoint jet algorithm used in this measurement
is infrared unsafe, as divergences appear in a fixed-order calculation
for configurations with three hard particles close in phase space
plus a soft one, as discussed in
Refs.~\cite{Salam:2009jx,Salam:2007xv}.
In order to compare the measured cross sections with a fixed-order
prediction, an infrared and collinear safe jet algorithm that is as
similar as possible to the midpoint algorithm, is used in the
prediction. This is the SISCone algorithm with the same
split-merge threshold of 0.75 and the same jet radius
$R=0.7$ of the midpoint algorithm used for the measured cross
sections. The additional uncertainty coming from the use of different
jet algorithms between data and theory is estimated by comparing the
particle-level cross sections for the two jet
algorithms. Figure~\ref{fig:CC_Pt1_Incl_JETALG} shows the cross
section ratios
of midpoint and SISCone jet algorithms for inclusive jet
\pt{} and rapidity in the \Zonejet{} final state. The difference at parton
level between SISCone and midpoint is between $2\%$ and
$3\%$. Larger differences between midpoint and SISCone are observed
if the underlying event is simulated; however, they do not affect the
comparison with fixed-order predictions. Figure~\ref{fig:CC_NJ_Incl_JETALG} shows the same
comparison as a function of jet multiplicity. The
difference at parton level between midpoint and SISCone is always
below $3\%$ and generally uniform.

\section{\label{sec:results}Results}
The differential cross sections of \Zjets{} production in \ppbar{}
collisions are measured independently in the
\Zee{} and \Zmm{} decay channels and combined using the best
linear unbiased estimate (BLUE) method~\cite{Lyons:1988rp}. 
The BLUE algorithm returns a weighted average of the measurements taking into
account different types of uncertainty and their
correlations. Systematic uncertainties related to trigger
efficiencies, lepton reconstruction efficiencies, and QCD and \Wjets{}
background  estimation are considered uncorrelated between the two
channels; all other contributions are treated as fully correlated. 

Inclusive \ZNjets{} cross sections are measured for
number of jets $N_{\textrm{jets}} \geqslant 1,2,3$, and $4$, various
differential cross sections are measured in the \Zonejet{},
\Ztwojets{}, and \Zthreejets{} final states.
Table \ref{tab:results} summarizes the measured cross sections.
\begin{table*}
    \caption{Summary of measured cross sections for each \ZNjets{} final state.
 }
  \begin{center}
    \label{tab:results}
    \begin{tabular}{ll}
      \toprule
      Final state              & Measured quantity~(Fig.) \\
      \colrule
      \ZNjets{}                & Inclusive cross section for $N_{\textrm{jets}} \geqslant 1,2,3$, and $4$~(\ref{fig:CC_NJ_Incl_APLB})  \\
      \Zonejet{}               & Leading jet \pt{}~(\ref{fig:CC_Pt1_Lead_APBEL}), 
                                 inclusive jet \pt{}~(\ref{fig:CC_Pt1_Incl_APML},\ref{fig:CC_Pt1_Incl_MCFM}), 
                                 inclusive jet $y$~(\ref{fig:CC_Y1_Incl_APBL},\ref{fig:CC_Y1_Incl_MCFM}), 
                                 $\pt^Z$~(\ref{fig:CC_ZPt1_APMEL}), $\Delta{\phi}_{Z,\textrm{jet}}$~(\ref{fig:CC_Zj_DPhi_LPA}),
                                 \htjet{}~(\ref{fig:CC_Htj1_MAPL})\\
      \Ztwojets{}              & $2\textit{nd}$ leading jet \pt{}~(\ref{fig:CC_Pt2_Lead}), inclusive-jet $y$~(\ref{fig:CC_Y2_Incl}),
                                 $M_{\mathit{jj}}$~(\ref{fig:CC_DJ_Mass_AM}), dijet $\Delta{R}$~(\ref{fig:CC_DJ_DR_AM}),
                                 dijet $\Delta{\phi}$~(\ref{fig:CC_DJ_DPhi_AM}), dijet $\Delta{y}$~(\ref{fig:CC_DJ_DY_AM}),
                                 $\theta_{Z,{\mathit{jj}}}$~(\ref{fig:CC_Zjj_Theta_AM})\\
      \Zthreejets{}            & $3$rd leading jet \pt{}~(\ref{fig:CC_Pt3_Y3_BH} a), inclusive-jet $y$~(\ref{fig:CC_Pt3_Y3_BH} b) \\
      \botrule
    \end{tabular}
  \end{center}
\end{table*}

\subsection{Cross section for the production of a \Zg{} boson in association with $N$ or more jets\label{sec:Znjet_results}}
\begin{figure*}
  \begin{center}
    \includegraphics[width=\figsizestar]{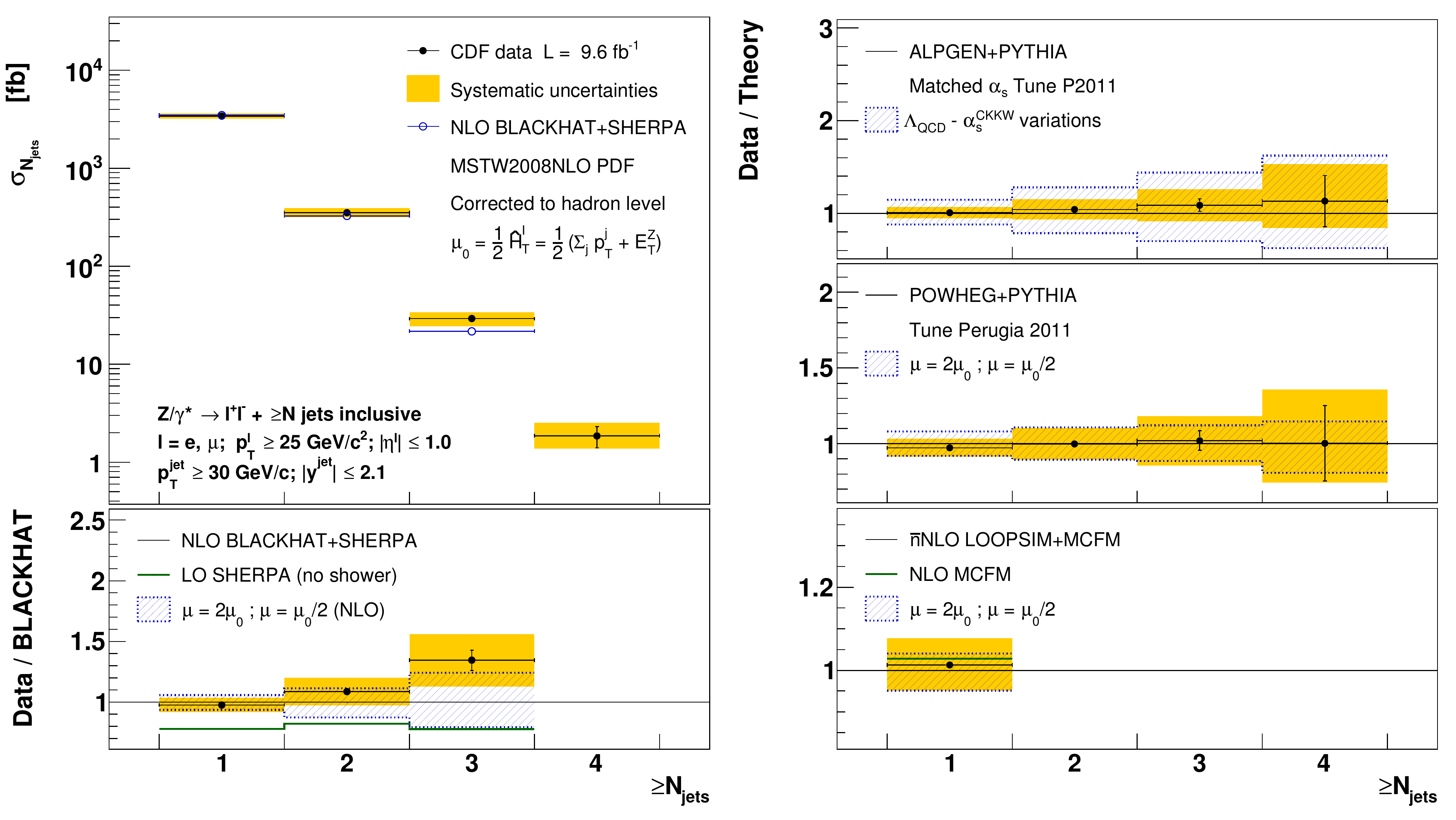}
    \caption{Inclusive \Zg{} + $\geqslant N$ jets cross section as a function
      of jet multiplicity.
      The measured cross section (black dots) is compared to the \blackhatsherpa{} NLO prediction (open circles).
      The black vertical bars show the statistical uncertainty, and
      the yellow bands show the total systematic uncertainty, except
      for the $5.8\%$ uncertainty on the luminosity.
      The lower and right panels show the data-to-theory ratio with respect to other theoretical predictions,
      with the blue dashed bands showing the scale-variation uncertainty of each prediction, which is associated with the 
      variation of the renormalization and factorization scales $\mu$ or to
      the combined variation of $\alpha_{s}^{\textrm{CKKW}}$ and $\Lambda_{QCD}$.
      \label{fig:CC_NJ_Incl_APLB}}
  \end{center}
\end{figure*}

The \ZNjets{} production cross sections are measured
for $N_{\textrm{jets}}$ up to four and compared to LO and NLO perturbative QCD
\blackhatsherpa{}, LO-ME+PS \alpgenpythia{}, and NLO+PS
\powhegpythia{} predictions. The \Zonejet{} cross section is compared
also to the \nnlo{} \loopsimmcfm{}
prediction. Figure~\ref{fig:CC_NJ_Incl_APLB} shows the inclusive cross
section as a function of jet multiplicity for \Zg{} + $\geqslant$ 1,
2, 3 and 4 jets. The measured cross section is in general good
agreement with all the predictions. The blue dashed bands show the
theoretical uncertainty associated with the variation of the
renormalization and factorization scales, except for the
\alpgenpythia{} prediction, where the band shows the uncertainty
associated with the variation of the CKKW renormalization scale. The
\alpgenpythia{} LO-ME+PS prediction provides a good model of the
measured cross sections, but has large theoretical uncertainty at
higher jet multiplicities. The \blackhatsherpa{} NLO perturbative QCD
prediction shows a reduced scale dependence with respect to the
\alpgenpythia{} LO-ME+PS prediction.
The \powhegpythia{} NLO+PS prediction has NLO accuracy only for
\Zonejet, but it can be compared to data in all the measured jet
multiplicities, where a general good agreement is observed.
The \loopsimmcfm{} \nnlo{} prediction is currently available only for
\Zonejet, where it shows a very good agreement with the measured cross
section and a reduced scale-variation uncertainty at the level of $5\%$.

The \Zthreejets{} \blackhatsherpa{} NLO perturbative QCD calculation
appears to be approximately $30\%$ lower than data, with the difference covered
by the scale-variation uncertainty. Such a difference is not observed in
the comparison with LO-ME+PS \alpgenpythia{} and NLO+PS
\powhegpythia{} predictions, 
in agreement with recent measurements using the anti-$k_t$ jet
algorithm~\cite{Aad:2013ysa, *Aad:2011qv}, which do not show any difference
with the NLO predictions at high jet multiplicities. The reason of
this difference has been found to be related to the different $\Delta
R$ angular reach~\cite{Salam:2009jx} between the SISCone and anti-$k_t$
algorithms, and how it is influenced by additional radiation between two hard
particles~\cite{Camarda:2012yha}. The difference between data or LO-ME+PS with respect to
the NLO prediction in the \Zthreejets{} final state is explained with
the presence of higher-order QCD radiation, which reduces the angular
reach of the SISCone algorithm and increases the cross section in this
particular configuration.

\subsection{Cross section for the production of a \Zg{} boson in
  association with one or more jets\label{sec:Z1jet_results}}
\begin{figure*}
  \begin{center}
    \includegraphics[width=\figsizestar]{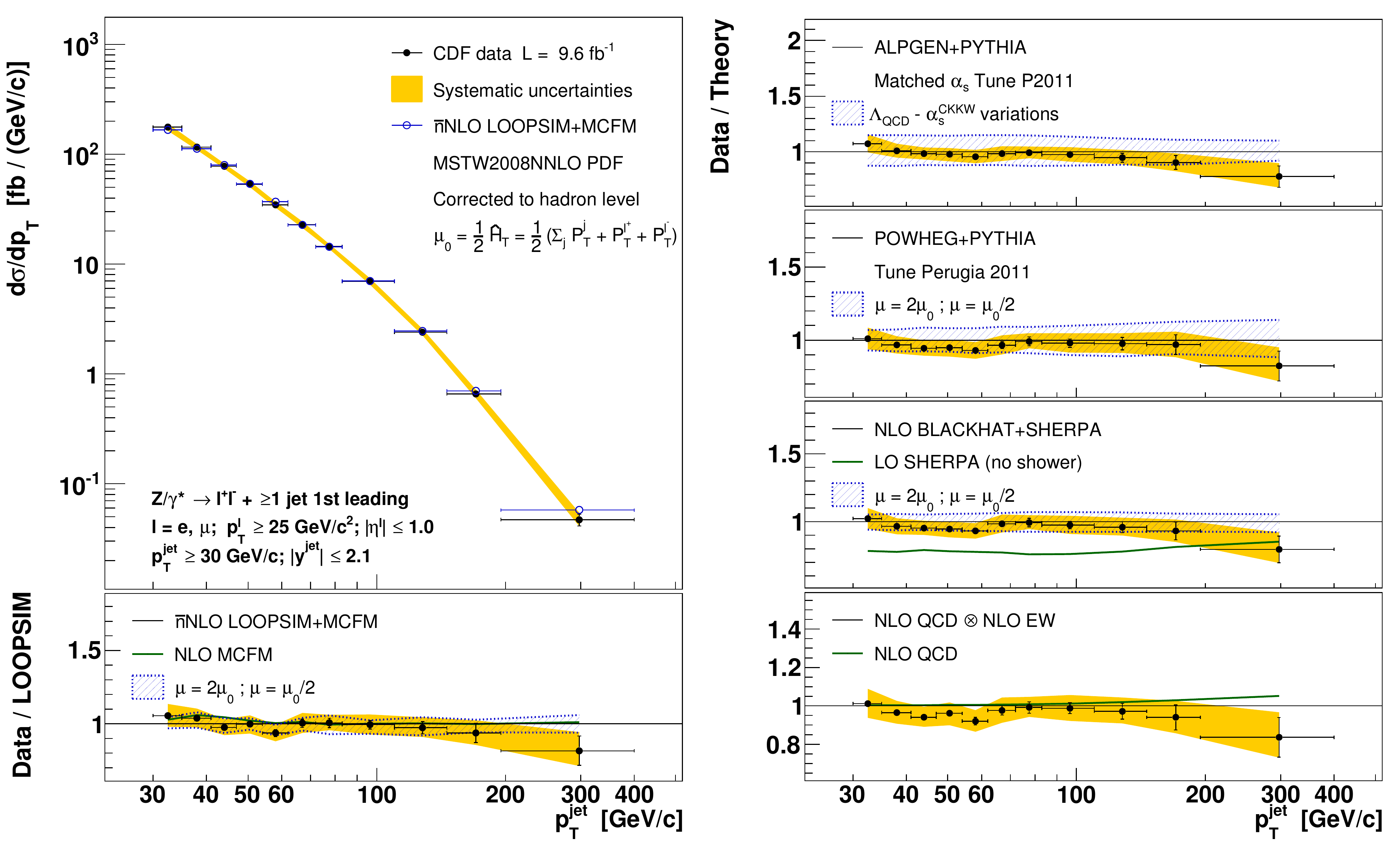}
    \caption{Differential cross section as a function of leading jet
      \pt{} for \Zonejet{} events.
      The measured cross section (black dots) is compared to the \loopsimmcfm{} \nnlo{} prediction (open circles).
      The black vertical bars show the statistical uncertainty, and
      the yellow bands show the total systematic uncertainty, except
      for the $5.8\%$ uncertainty on the luminosity.
      The lower and right panels show the data-to-theory ratio with respect to other theoretical predictions,
      with the blue dashed bands showing the scale-variation uncertainty of each prediction, which is associated with the 
      variation of the renormalization and factorization scales $\mu$ or to
      the combined variation of $\alpha_{s}^{\textrm{CKKW}}$ and $\Lambda_{QCD}$.
      \label{fig:CC_Pt1_Lead_APBEL}}
  \end{center}
\end{figure*}

\begin{figure*}
  \begin{center}
    \includegraphics[width=\figsizestar]{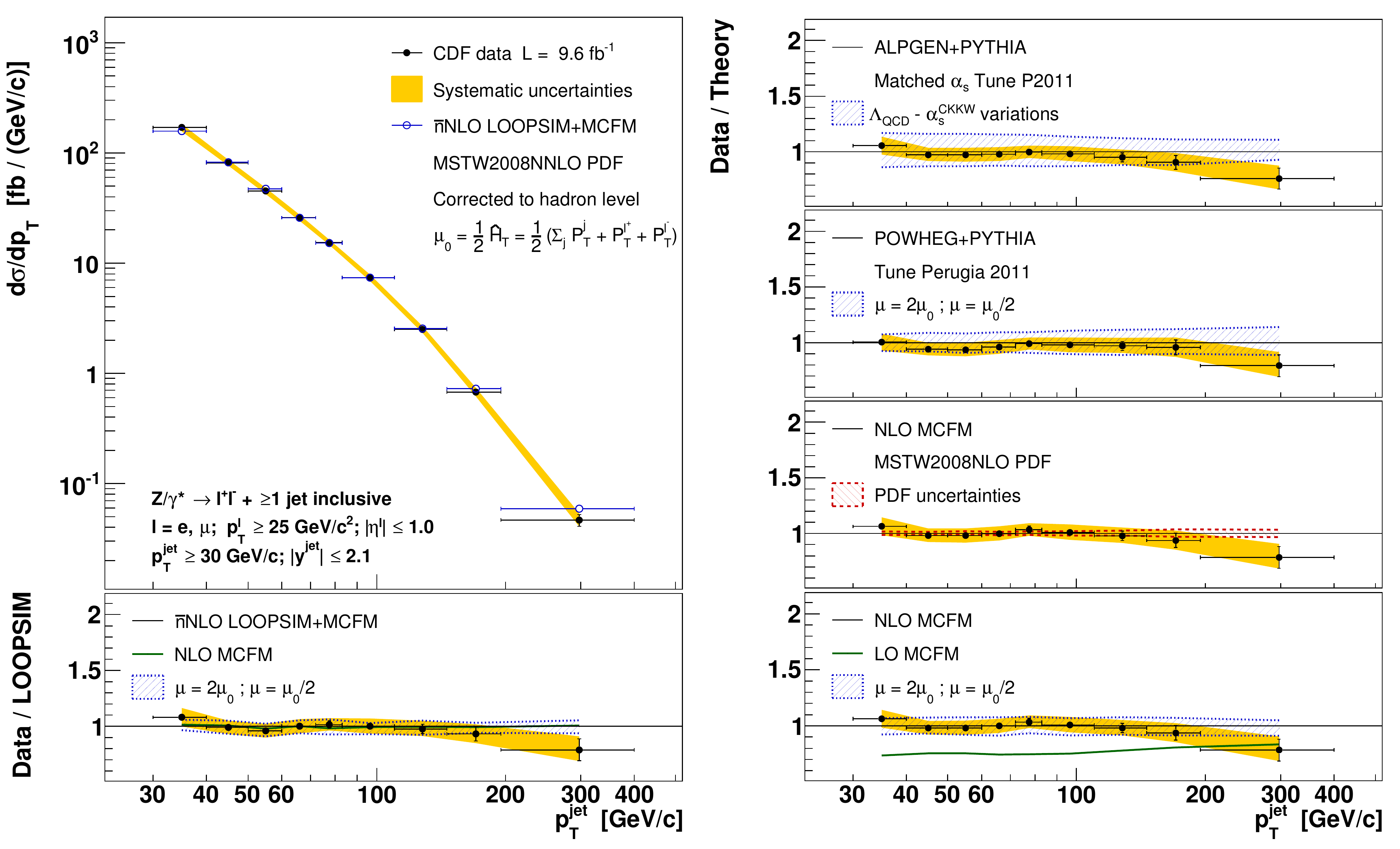}
    \caption{Differential cross section as a function of inclusive jet
      \pt{} for \Zonejet{} events.
      The measured cross section (black dots) is compared to the \loopsimmcfm{} \nnlo{} prediction (open circles).
      The black vertical bars show the statistical uncertainty, and
      the yellow bands show the total systematic uncertainty, except
      for the $5.8\%$ uncertainty on the luminosity.
      The lower and right panels show the data-to-theory ratio with
      respect to other theoretical predictions with the blue dashed
      bands showing the scale-variation uncertainty of each
      prediction, which is associated with the variation of the
      renormalization and factorization scales $\mu$ or to the
      combined variation of $\alpha_{s}^{\textrm{CKKW}}$ and
      $\Lambda_{QCD}$. The red dashed band shows the PDF uncertainty
      evaluated with the \mcfm{} prediction.
      \label{fig:CC_Pt1_Incl_APML}}
  \end{center}
\end{figure*}

\begin{figure*}
  \begin{center}
    \includegraphics[width=\figsizestar]{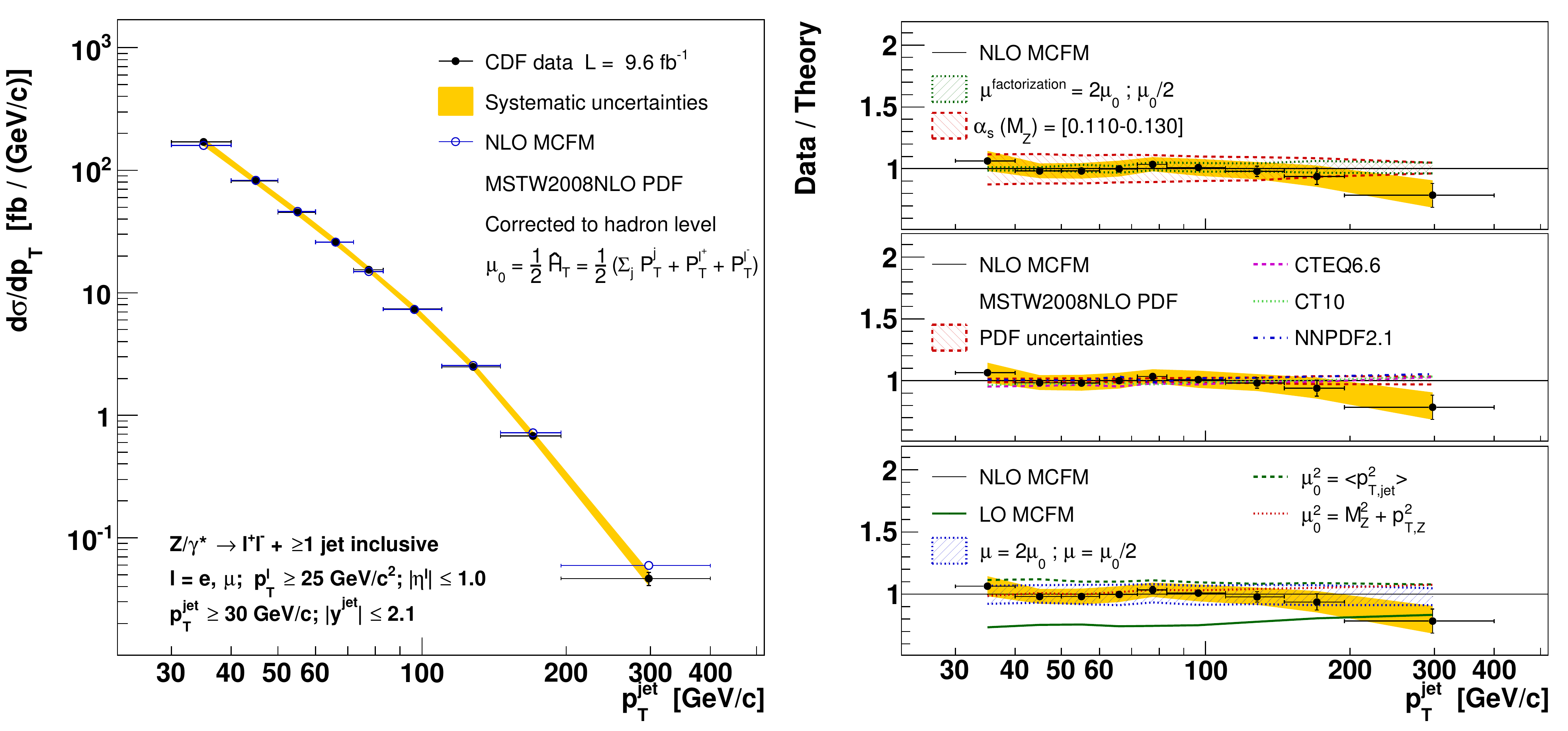}
    \caption{Differential cross section as a function of inclusive jet
      \pt{} for \Zonejet{} events.
      The measured cross section (black dots) is compared to the \mcfm{} NLO prediction (open circles).
      The black vertical bars show the statistical uncertainty, and
      the yellow bands show the total systematic uncertainty, except
      for the $5.8\%$ uncertainty on the luminosity.
      The right panels show, from top to bottom, the data-to-theory
      ratio including variations of $\alpha_s(M_Z)$ (red dashed band) and factorization
      scale (green dashed band); various PDF sets and PDF
      uncertainty (red dashed band); and various choice of the functional form
      of the factorization and renormalization scales and
      scale-variation uncertainty (blue dashed band).
      \label{fig:CC_Pt1_Incl_MCFM}}
  \end{center}
\end{figure*}

Figures~\ref{fig:CC_Pt1_Lead_APBEL} and~\ref{fig:CC_Pt1_Incl_APML}
show the leading-jet and inclusive-jet cross sections differential in
\pt{} for \Zonejet{} events.
All the theoretical predictions are in
reasonable agreement with the measured cross sections. The NLO
electroweak corrections give a $5\%$ negative contribution in the last
\Zg{} and leading jet \pt{} bin, due to the large Sudakov
logarithms that appear in the virtual part of the calculation~\cite{Denner:2011vu}.
The scale-variation uncertainty is quite independent of the jet \pt{} and of the
order of $4\%-6\%$ for the \nnlo{} \loopsim{}
prediction. Figure~\ref{fig:CC_Pt1_Incl_MCFM} shows variations in the
\mcfm{} prediction with different values of the strong-interaction coupling
constant at the $Z$ boson mass, $\alpha_s(M_Z)$, factorization scale,
PDF sets, and choice of the functional form of the factorization and renormalization scales.

\begin{figure*}
  \begin{center}
    \includegraphics[width=\figsizestar]{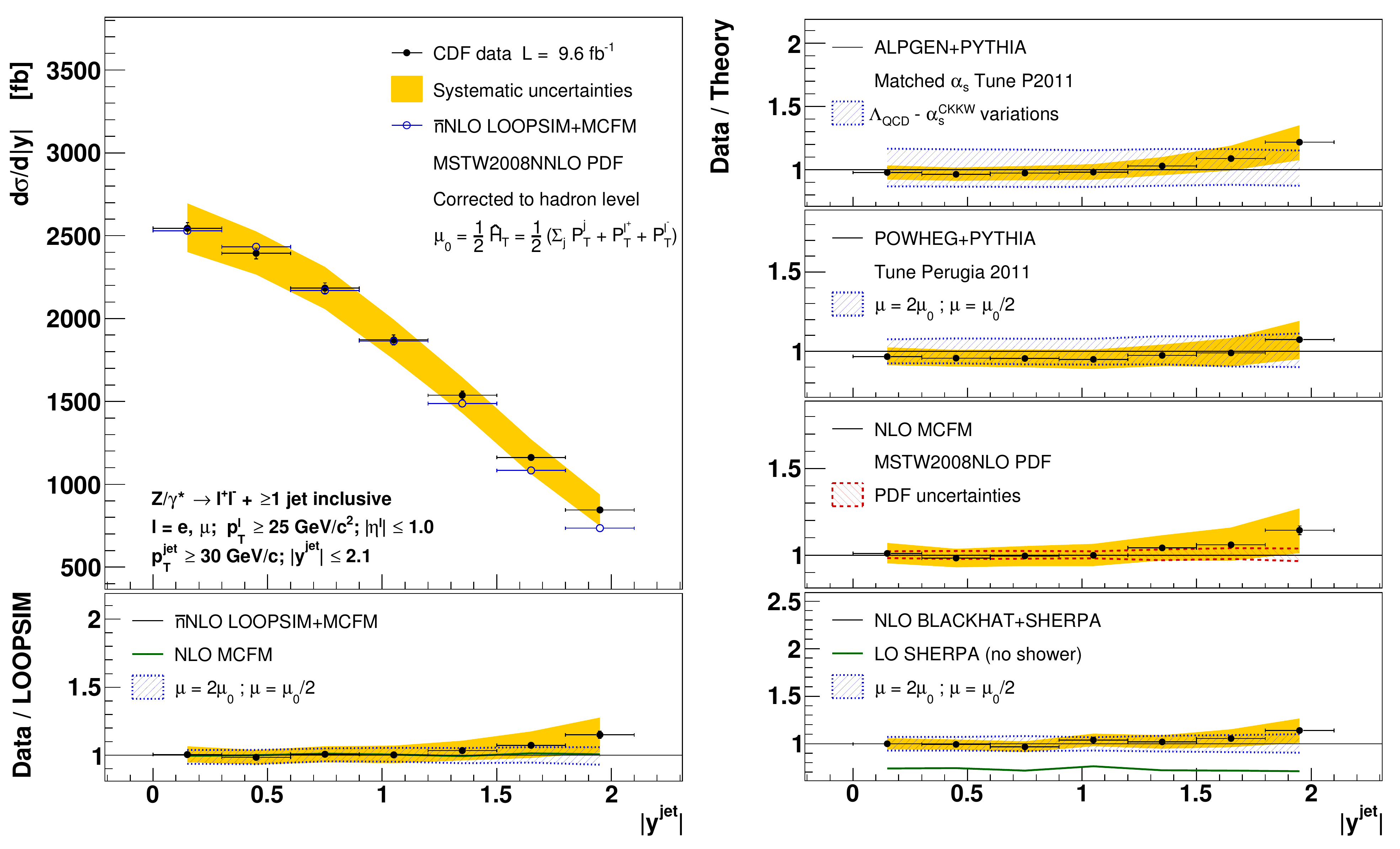}
    \caption{Differential cross section as a function
      of inclusive jet rapidity for \Zonejet{} events.
      The measured cross section (black dots) is compared to the \loopsimmcfm{} \nnlo{} prediction (open circles).
      The black vertical bars show the statistical uncertainty, and
      the yellow bands show the total systematic uncertainty, except
      for the $5.8\%$ uncertainty on the luminosity.
      The lower and right panels show the data-to-theory ratio with respect to other theoretical predictions,
      with the blue dashed bands showing the scale-variation uncertainty of each prediction, which is associated with the 
      variation of the renormalization and factorization scales $\mu$ or to
      the combined variation of $\alpha_{s}^{\textrm{CKKW}}$ and
      $\Lambda_{QCD}$.
      The red dashed band shows the PDF uncertainty
      evaluated with the \mcfm{} prediction.
      \label{fig:CC_Y1_Incl_APBL}}
  \end{center}
\end{figure*}

\begin{figure*}
  \begin{center}
    \includegraphics[width=\figsizestar]{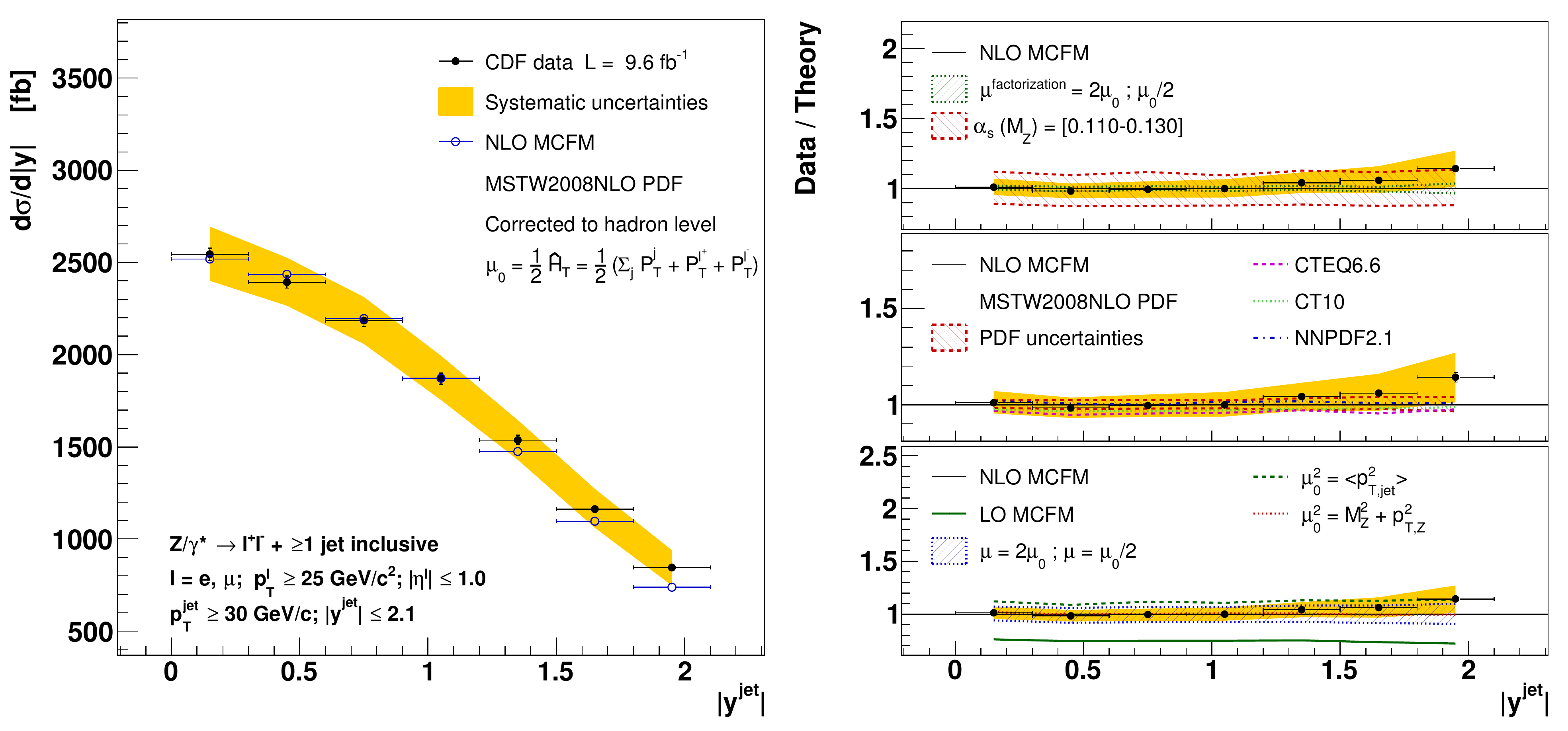}
    \caption{Differential cross section as a function
      of inclusive jet rapidity for \Zonejet{} events.
      The measured cross section (black dots) is compared to the \mcfm{} NLO prediction (open circles).
      The black vertical bars show the statistical uncertainty, and
      the yellow bands show the total systematic uncertainty, except
      for the $5.8\%$ uncertainty on the luminosity.
      The right panels show, from top to bottom, the data-to-theory
      ratio including variations of $\alpha_s(M_Z)$ (red dashed band) and factorization
      scale (green dashed band); various PDF sets and PDF
      uncertainty (red dashed band); and various choice of the functional form
      of the factorization and renormalization scales and
      scale-variation uncertainty (blue dashed band).
      \label{fig:CC_Y1_Incl_MCFM}}
  \end{center}
\end{figure*}

Figure~\ref{fig:CC_Y1_Incl_APBL} shows the inclusive-jet cross
sections differential in rapidity for \Zonejet{} events. All predictions
correctly model this quantity. In the high-rapidity region the
measured cross section is higher than predictions; however, the
difference is covered by the uncertainty due to the contribution of 
multiple \ppbar{} interaction. The \nnlo{} \loopsimmcfm{} prediction
has the lowest scale-variation theoretical uncertainty, which is of the order of $4\%-6\%$,
and the PDF uncertainty is between $2\%$ and $4\%$.
In the high-rapidity region the \alpgen{} prediction is lower than other
theoretical models; however, the difference with data is covered by the
large CKKW renormalization scale-variation uncertainty of this prediction.
Figure~\ref{fig:CC_Y1_Incl_MCFM} shows variations in the
\mcfm{} prediction with different values of $\alpha_s(M_Z)$, factorization scale,
PDF sets, and choice of the functional form of the factorization and renormalization scales.

\begin{figure*}
  \begin{center}
    \includegraphics[width=\figsizestar]{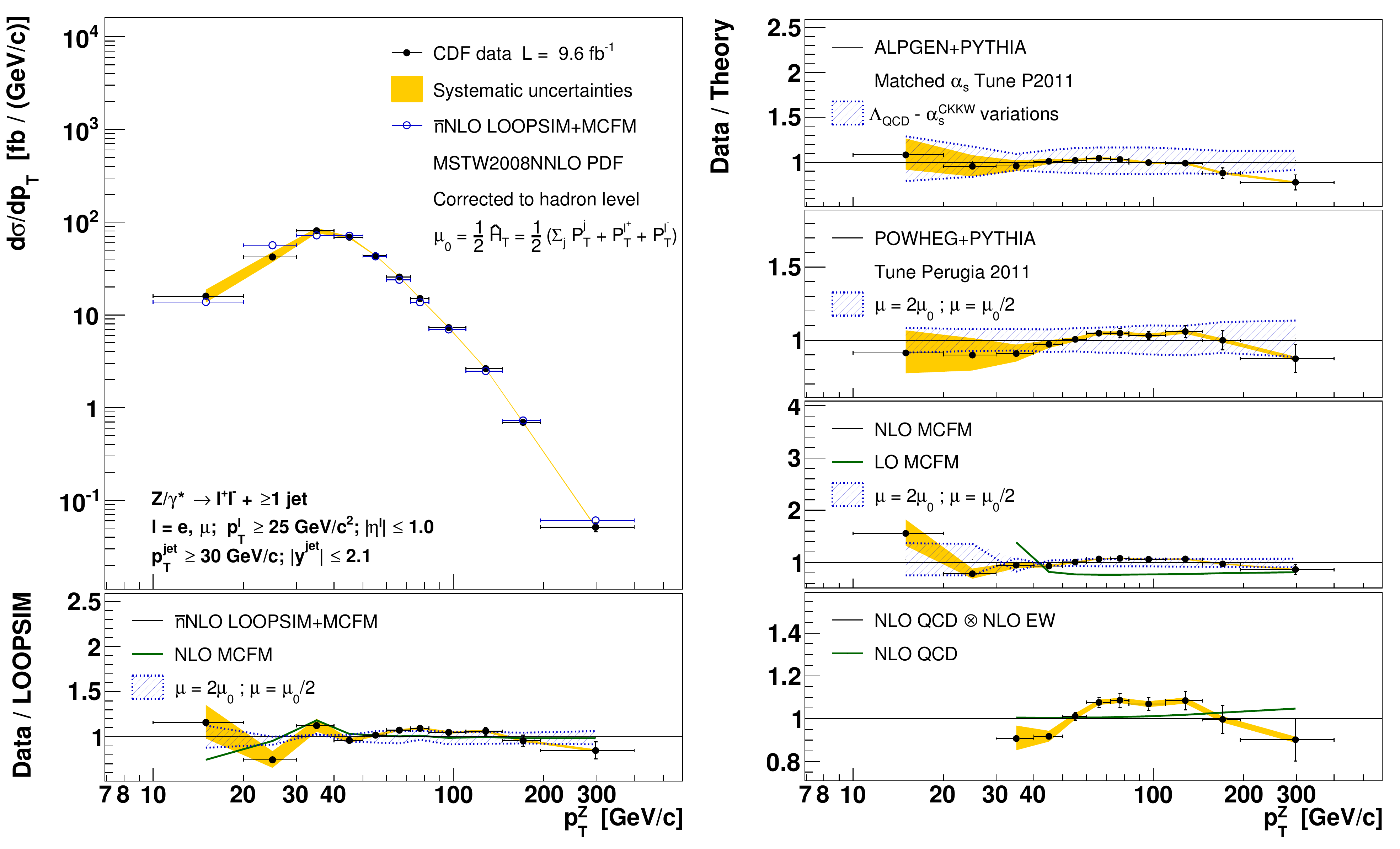}
    \caption{Differential cross section as a function
      of \Zg{} \pt{} for \Zonejet{} events.
      The measured cross section (black dots) is compared to the \loopsimmcfm{} \nnlo{} prediction (open circles).
      The black vertical bars show the statistical uncertainty, and
      the yellow bands show the total systematic uncertainty, except
      for the $5.8\%$ uncertainty on the luminosity.
      The lower and right panels show the data-to-theory ratio with respect to other theoretical predictions,
      with the blue dashed bands showing the scale-variation uncertainty of each prediction, which is associated with the 
      variation of the renormalization and factorization scales $\mu$ or to
      the combined variation of $\alpha_{s}^{\textrm{CKKW}}$ and $\Lambda_{QCD}$.
      \label{fig:CC_ZPt1_APMEL}}
  \end{center}
\end{figure*}

Figure~\ref{fig:CC_ZPt1_APMEL} shows the production cross section
differential in $\pt(\Zg{})$ for the \Zonejet{} final state. The perturbative QCD
fixed-order calculations \mcfm{} and \loopsimmcfm{} fail in describing
the region below the $30$~\gevc{} jet \pt{} threshold, where
multiple-jet emission and nonperturbative QCD corrections are
significant. The low \Zg{} \pt{} region is better described by the
\alpgenpythia{} and \powhegpythia{} predictions, which include parton
shower radiation, and in which the nonperturbative QCD corrections
are applied as part of the \pythia{} MC event evolution. In
the intermediate \Zg{} \pt{} region, the ratios of the data over the NLO
\mcfm{}, NLO+PS \powhegpythia{} and \nnlo{} \loopsimmcfm{} predictions
show a slightly concave shape, which is covered by the scale-variation 
uncertainty. The NLO electroweak corrections
related to the large Sudakov logarithms are negative and of the order
of $5\%$ in the last \pt{} bin.

\begin{figure*}
  \begin{center}
    \includegraphics[width=\figsizestar]{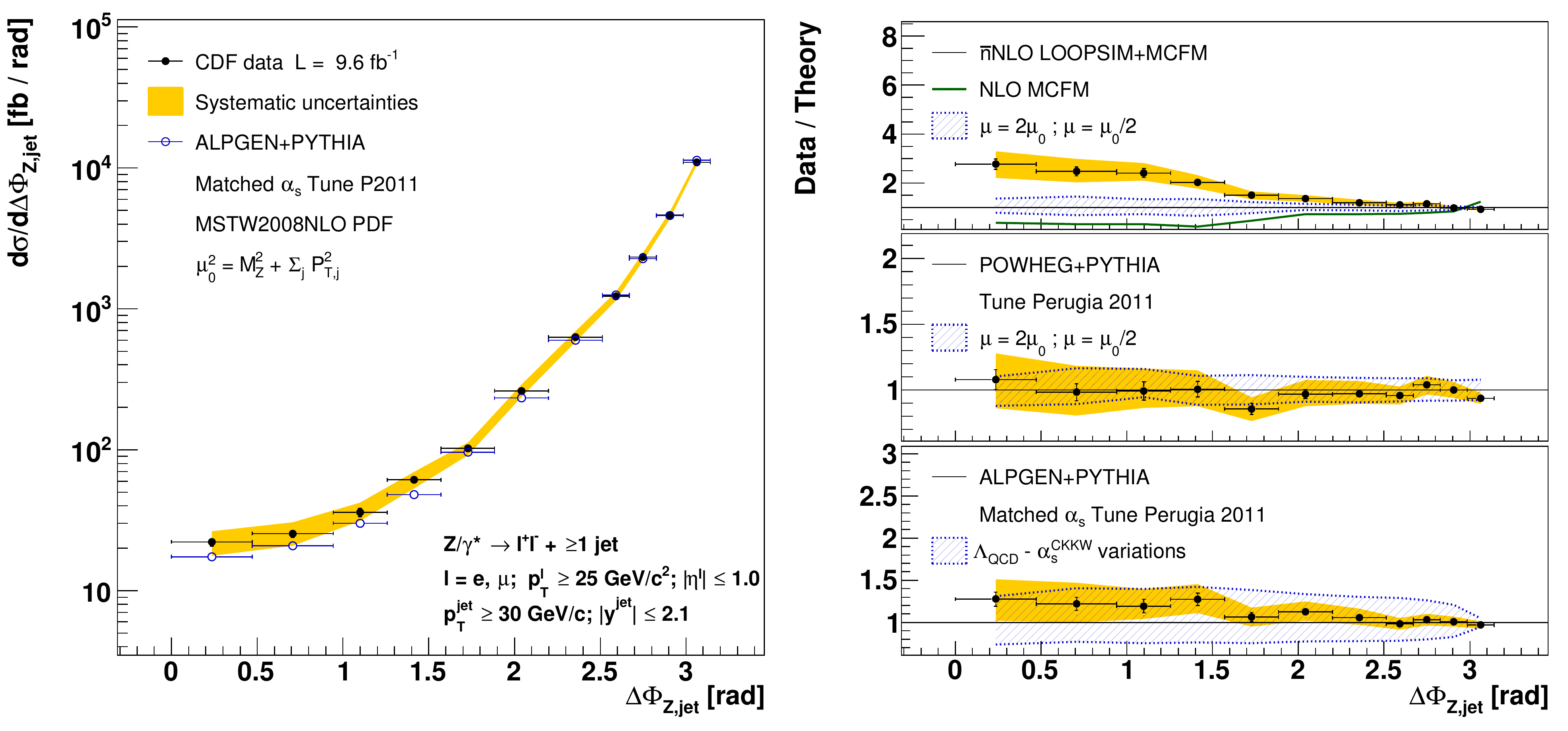}
    \caption{Differential cross section as a function
      of \Zg{}-jet $\Delta{\phi}$ for \Zonejet{} events.
      The measured cross section (black dots) is compared to the \alpgenpythia{} prediction (open circles).
      The black vertical bars show the statistical uncertainty, and
      the yellow bands show the total systematic uncertainty, except
      for the $5.8\%$ uncertainty on the luminosity.
      The right panels show the data-to-theory ratio with respect to other theoretical predictions,
      with the blue dashed bands showing the scale-variation uncertainty of each prediction, which is associated with the 
      variation of the renormalization and factorization scales $\mu$ or to
      the combined variation of $\alpha_{s}^{\textrm{CKKW}}$ and $\Lambda_{QCD}$.
      \label{fig:CC_Zj_DPhi_LPA}}
  \end{center}
\end{figure*}

Figure~\ref{fig:CC_Zj_DPhi_LPA} shows the differential cross section
as a function of the \Zg{}-leading jet $\Delta{\phi}$ variable in \Zonejet{}
events. The \alpgenpythia{} prediction shows good agreement with the measured cross
section in the region $\Delta{\phi} \geqslant \pi / 2$. In the region
$\Delta{\phi} < \pi / 2$ the \alpgenpythia{} prediction is lower
than the data, with the difference covered by the scale-variation
uncertainty. The \powhegpythia{} prediction has very good agreement
with the data over all of the \Zg{}-jet $\Delta{\phi}$ spectrum, and
is affected by smaller scale-variation uncertainty. The difference between the \alpgenpythia{} and
\powhegpythia{} predictions is comparable to the experimental
systematic uncertainty, which is dominated by the uncertainty from the
contribution of multiple \ppbar{} interactions. Hence, the measured
cross section cannot be used to distinguish between the two
models. The NLO \mcfm{} prediction fails to describe the region
$\Delta{\phi} < \pi / 2$ because it does not include the \Zg{} + 3 jets configuration, whereas
\nnlo{} \loopsimmcfm{}, which includes the \Zg{} + 3 jets with only LO
accuracy, predicts a rate approximately 2--3 times smaller than the
rate observed in data in this region.

\begin{figure*}
  \begin{center}
    \includegraphics[width=\figsizestar]{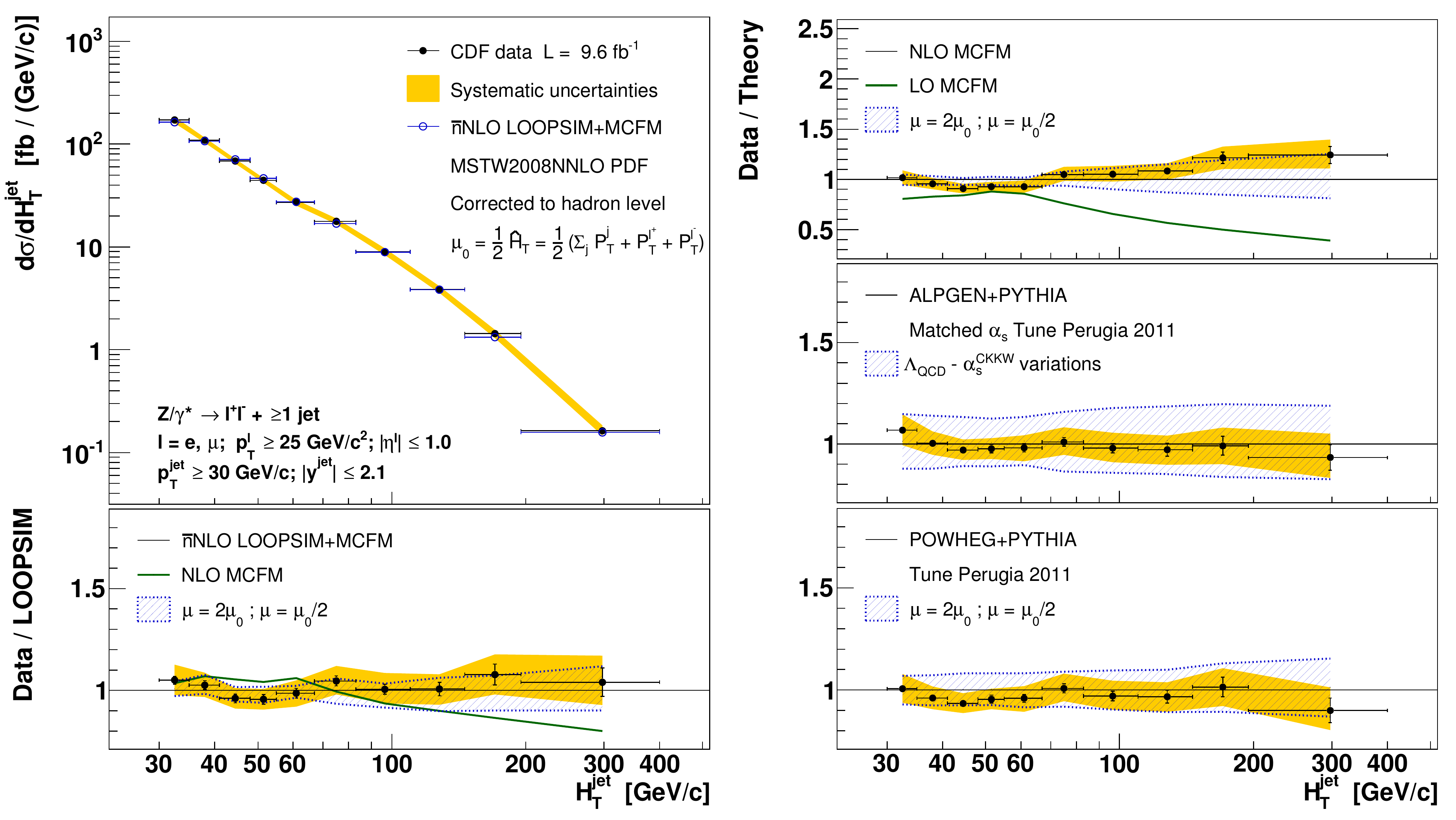}
    \caption{Differential cross section as a function
      of $\htjet = \sum{\ptjet}$ for \Zonejet{} events.
      The measured cross section (black dots) is compared to the \loopsimmcfm{} \nnlo{} prediction (open circles).
      The black vertical bars show the statistical uncertainty, and
      the yellow bands show the total systematic uncertainty, except
      for the $5.8\%$ uncertainty on the luminosity.
      The lower and right panels show the data-to-theory ratio with respect to other theoretical predictions,
      with the blue dashed bands showing the scale-variation uncertainty of each prediction, which is associated with the 
      variation of the renormalization and factorization scales $\mu$ or to
      the combined variation of $\alpha_{s}^{\textrm{CKKW}}$ and $\Lambda_{QCD}$.
      \label{fig:CC_Htj1_MAPL}}
  \end{center}
\end{figure*}

Some \Zjets{} observables have larger NLO-to-LO K-factors, defined as
the ratio of the NLO prediction over the LO prediction, and are expected
to have significant corrections at higher order than NLO~\cite{Rubin:2010xp}.
The most remarkable example is the \htjet, defined as $\htjet = \sum{\ptjet}$, in
\Zonejet{} events. Figure~\ref{fig:CC_Htj1_MAPL} shows the measured
cross section as a function of \htjet{} compared to the available
theoretical predictions. The NLO \mcfm{} prediction fails to describe
the shape of the \htjet{} distribution, in particular it
underestimates the measured cross section in the high \htjet{} region,
where the NLO-to-LO K-factor is greater than approximately two and a
larger NLO scale-variation uncertainty is observed. The LO-ME+PS \alpgenpythia{} prediction is in
good agreement with data, but suffers for the large LO scale
uncertainty. The \powhegpythia{} prediction also is in good agreement with data,
but is still affected by the larger NLO scale-variation uncertainty in the high
\pt{} tail. The \nnlo{} \loopsimmcfm{} prediction provides a good
modeling of the data distribution, and shows a significantly reduced
scale-variation uncertainty.

\subsection{Cross section for the production of a \Zg{} boson in association with two or more jets \label{sec:Z2jet_results}}
\begin{figure*}
  \begin{center}
    \includegraphics[width=\figsizestar]{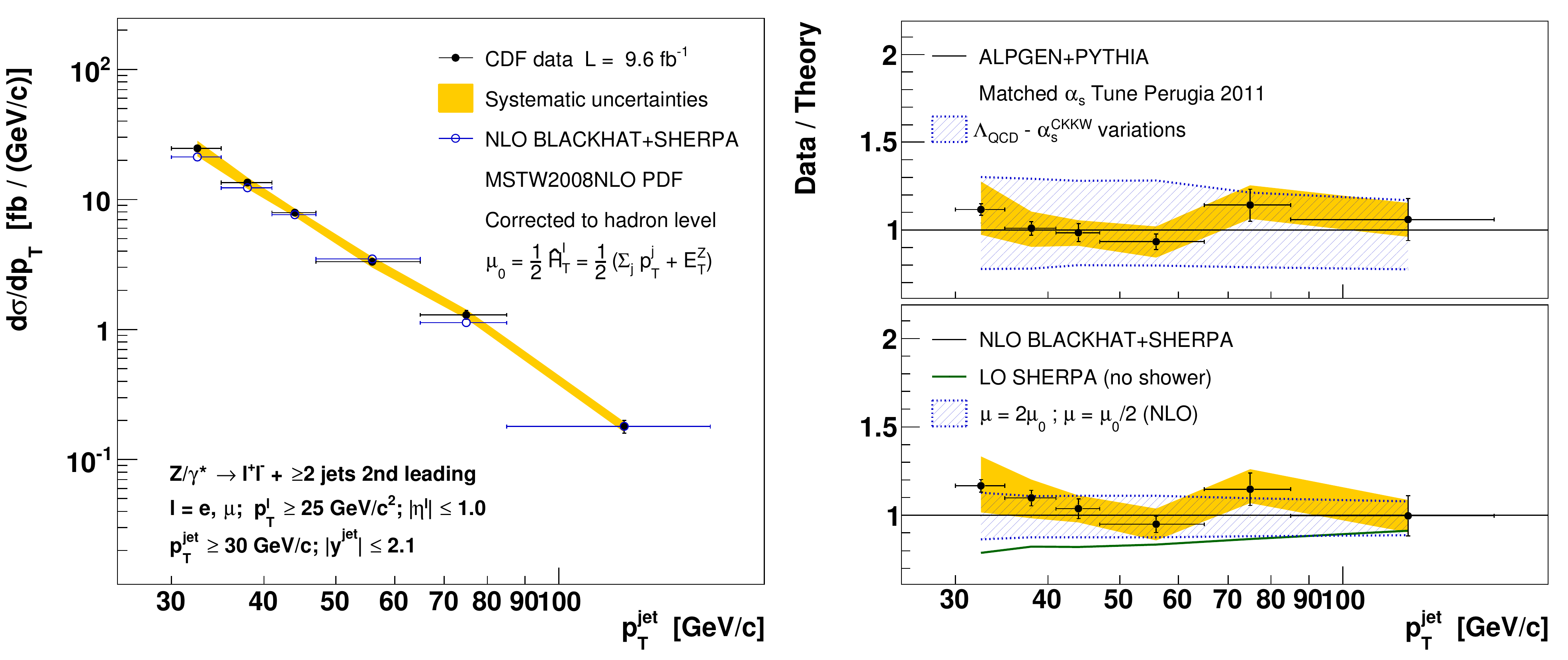}
    \caption{Differential cross section as a function
      of $2\textit{nd}$ leading jet \pt{} for \Ztwojets{} events.
      The measured cross section (black dots) is compared to the \blackhatsherpa{} NLO prediction (open circles).
      The black vertical bars show the statistical uncertainty, and
      the yellow bands show the total systematic uncertainty, except
      for the $5.8\%$ uncertainty on the luminosity. 
      The right panels show the data-to-theory ratio with respect to \alpgenpythia{} and \blackhatsherpa{} predictions,
      with the blue dashed bands showing the scale-variation uncertainty of each prediction, which is associated with the 
      variation of the renormalization and factorization scales $\mu$ or to
      the combined variation of $\alpha_{s}^{\textrm{CKKW}}$ and $\Lambda_{QCD}$.
      \label{fig:CC_Pt2_Lead}}
  \end{center}
\end{figure*}

\begin{figure*}
  \begin{center}
    \includegraphics[width=\figsizestar]{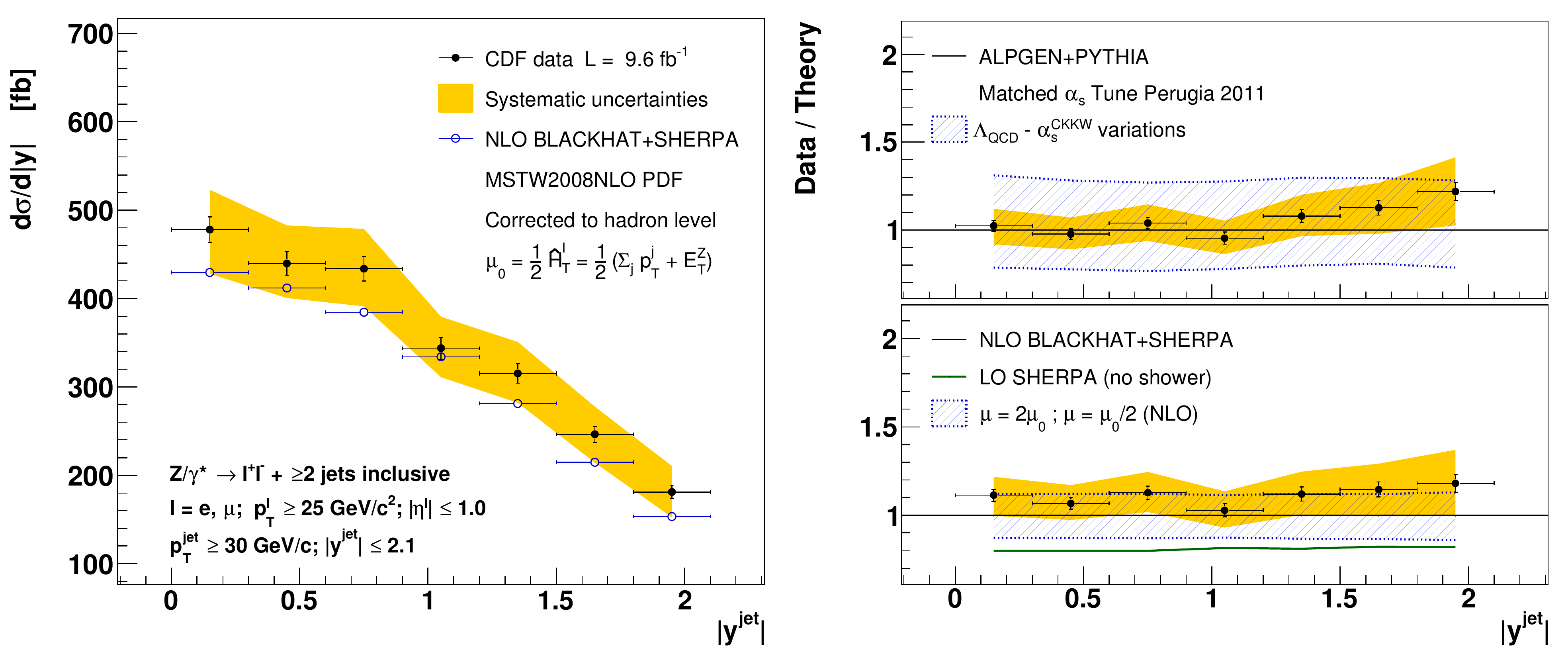}
    \caption{Differential cross section as a function
      of inclusive jet rapidity for \Ztwojets{} events.
      The measured cross section (black dots) is compared to the \blackhatsherpa{} NLO prediction (open circles).
      The black vertical bars show the statistical uncertainty, and
      the yellow bands show the total systematic uncertainty, except
      for the $5.8\%$ uncertainty on the luminosity.
      The right panels show the data-to-theory ratio with respect to \alpgenpythia{} and \blackhatsherpa{} predictions,
      with the blue dashed bands showing the scale-variation uncertainty of each prediction, which is associated with the 
      variation of the renormalization and factorization scales $\mu$ or to
      the combined variation of $\alpha_{s}^{\textrm{CKKW}}$ and $\Lambda_{QCD}$.
      \label{fig:CC_Y2_Incl}}
  \end{center}
\end{figure*}

\begin{figure*}
  \begin{center}
    \includegraphics[width=\figsizestar]{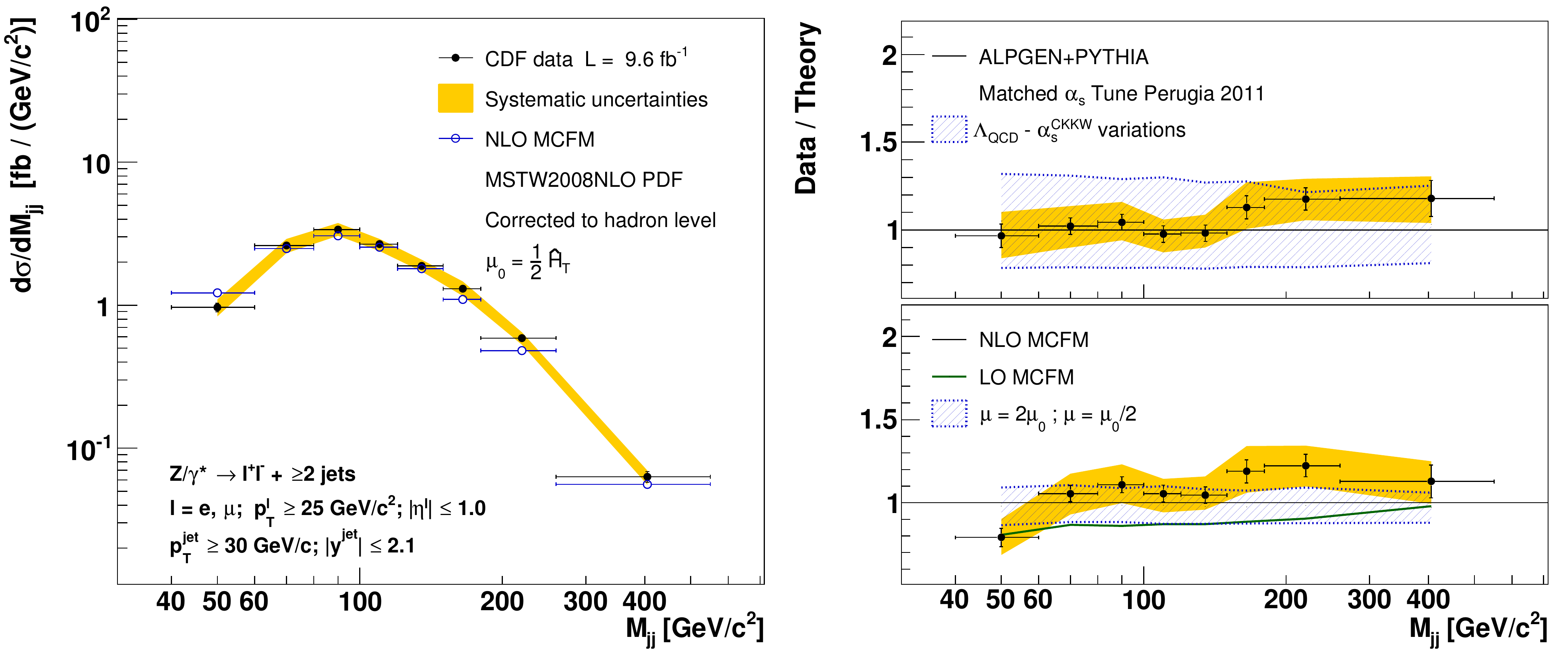}
    \caption{Differential cross section as a function
      of dijet mass $M_{\mathit{jj}}$ for \Ztwojets{} events.
      The measured cross section (black dots) is compared to the \mcfm{} NLO prediction (open circles).
      The black vertical bars show the statistical uncertainty, and
      the yellow bands show the total systematic uncertainty, except
      for the $5.8\%$ uncertainty on the luminosity.
      The right panels show the data-to-theory ratio with respect to \alpgenpythia{} and \mcfm{} predictions,
      with the blue dashed bands showing the scale-variation uncertainty of each prediction, which is associated with the 
      variation of the renormalization and factorization scales $\mu$ or to
      the combined variation of $\alpha_{s}^{\textrm{CKKW}}$ and $\Lambda_{QCD}$.
      \label{fig:CC_DJ_Mass_AM}}
  \end{center}
\end{figure*}

\begin{figure*}
  \begin{center}
    \includegraphics[width=\figsizestar]{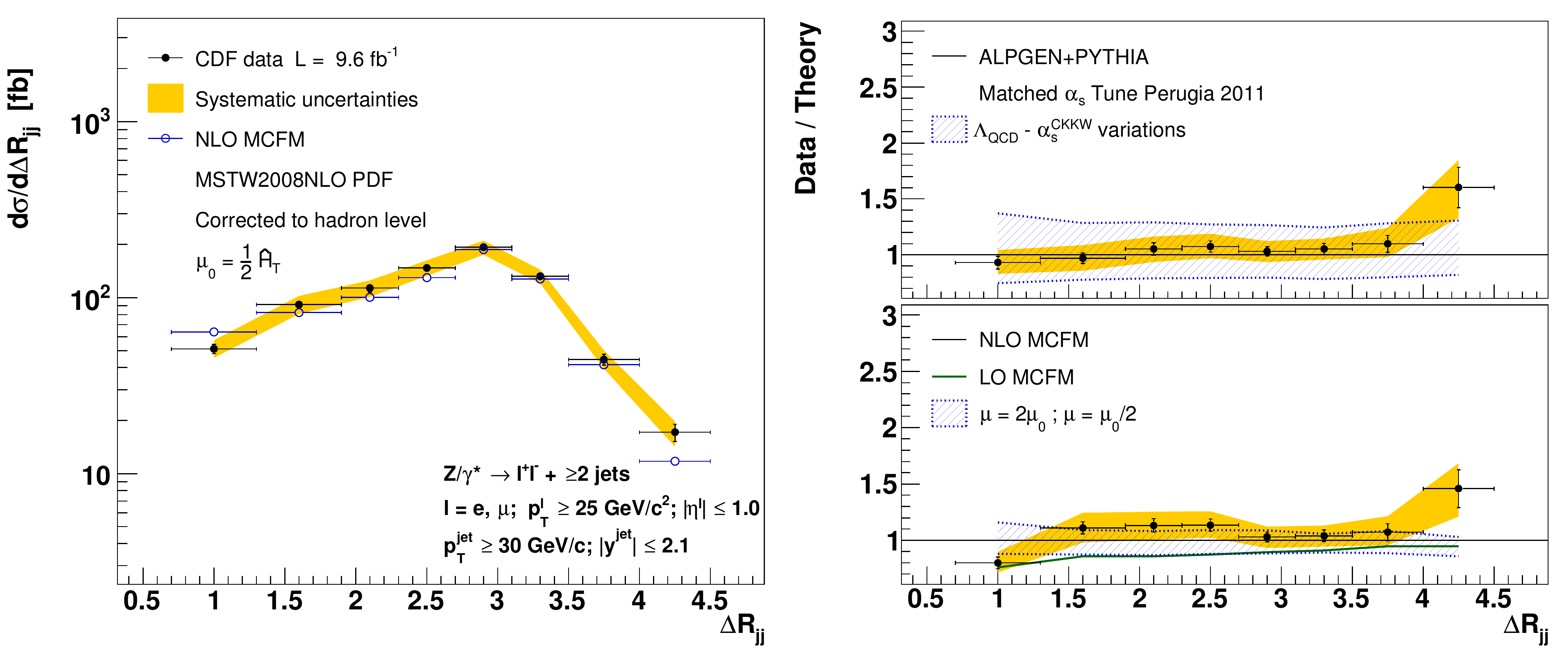}
    \caption{Differential cross section as a function
      of dijet $\Delta{R}$ for \Ztwojets{} events.
      The measured cross section (black dots) is compared to the \mcfm{} NLO prediction (open circles).
      The black vertical bars show the statistical uncertainty, and
      the yellow bands show the total systematic uncertainty, except
      for the $5.8\%$ uncertainty on the luminosity.
      The right panels show the data-to-theory ratio with respect to \alpgenpythia{} and \mcfm{} predictions,
      with the blue dashed bands showing the scale-variation uncertainty of each prediction, which is associated with the 
      variation of the renormalization and factorization scales $\mu$ or to
      the combined variation of $\alpha_{s}^{\textrm{CKKW}}$ and $\Lambda_{QCD}$.
      \label{fig:CC_DJ_DR_AM}}
  \end{center}
\end{figure*}

\begin{figure*}
  \begin{center}
    \includegraphics[width=\figsizestar]{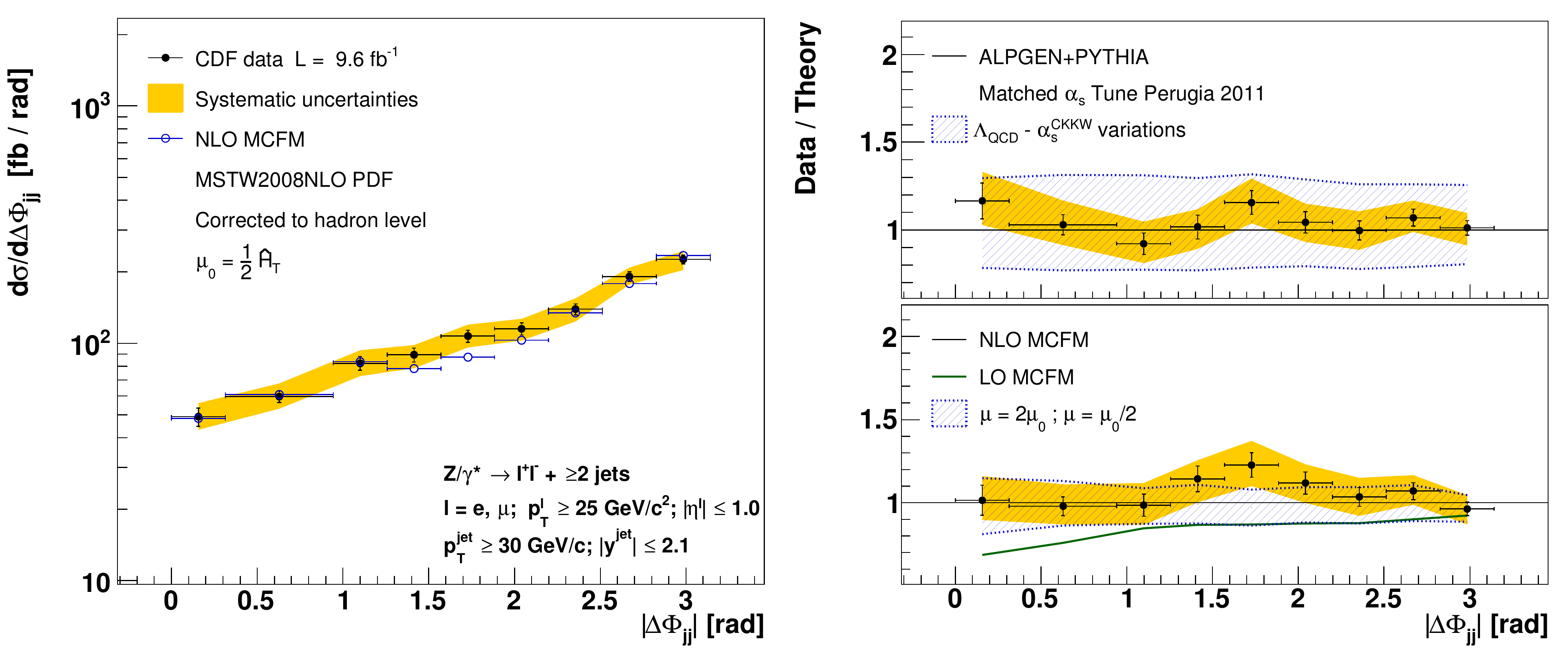}
    \caption{Differential cross section as a function
      of dijet $\Delta{\phi}$ for \Ztwojets{} events.
      The measured cross section (black dots) is compared to the \mcfm{} NLO prediction (open circles).
      The black vertical bars show the statistical uncertainty, and
      the yellow bands show the total systematic uncertainty, except
      for the $5.8\%$ uncertainty on the luminosity.
      The right panels show the data-to-theory ratio with respect to \alpgenpythia{} and \mcfm{} predictions,
      with the blue dashed bands showing the scale-variation uncertainty of each prediction, which is associated with the 
      variation of the renormalization and factorization scales $\mu$ or to
      the combined variation of $\alpha_{s}^{\textrm{CKKW}}$ and $\Lambda_{QCD}$.
      \label{fig:CC_DJ_DPhi_AM}}
  \end{center}
\end{figure*}

\begin{figure*}
  \begin{center}
    \includegraphics[width=\figsizestar]{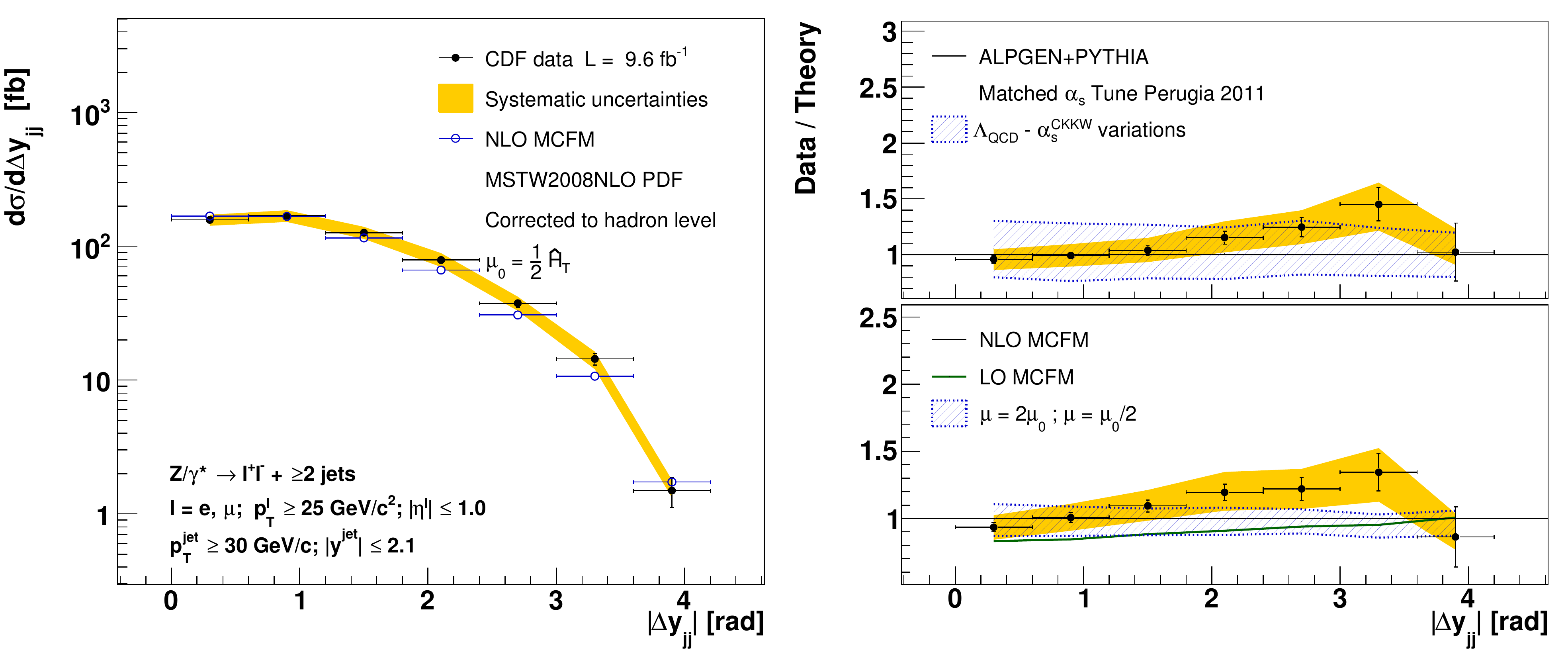}
    \caption{Differential cross section as a function
      of dijet $\Delta{y}$ for \Ztwojets{} events.
      The measured cross section (black dots) is compared to the \mcfm{} NLO prediction (open circles).
      The black vertical bars show the statistical uncertainty, and
      the yellow bands show the total systematic uncertainty, except
      for the $5.8\%$ uncertainty on the luminosity.
      The right panels show the data-to-theory ratio with respect to \alpgenpythia{} and \mcfm{} predictions,
      with the blue dashed bands showing the scale-variation uncertainty of each prediction, which is associated with the 
      variation of the renormalization and factorization scales $\mu$ or to
      the combined variation of $\alpha_{s}^{\textrm{CKKW}}$ and $\Lambda_{QCD}$.
      \label{fig:CC_DJ_DY_AM}}
  \end{center}
\end{figure*}

\begin{figure*}
  \begin{center}
    \includegraphics[width=\figsizestar]{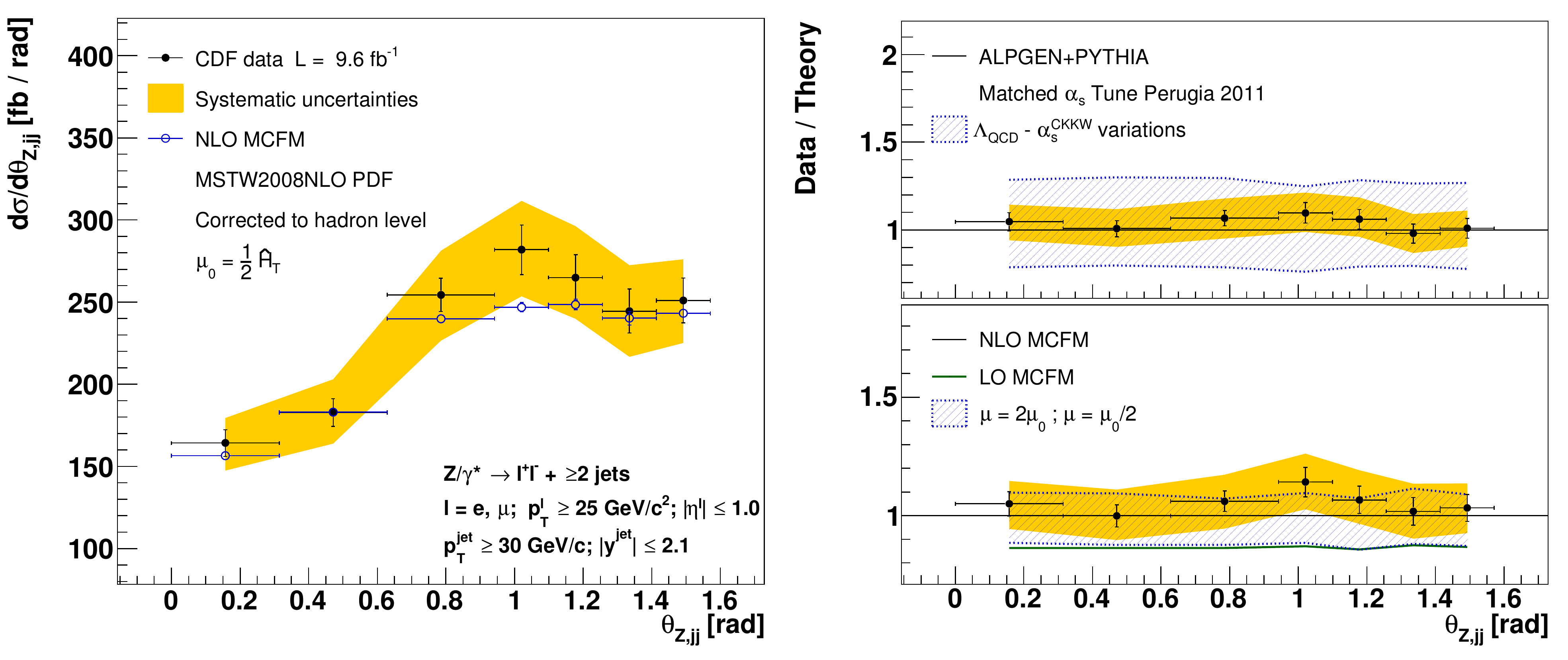}
    \caption{Differential cross section as a function
      of the dihedral angle $\theta_{Z,{\mathit{jj}}}$ for \Ztwojets{} events.
      The measured cross section (black dots) is compared to the \mcfm{} NLO prediction (open circles).
      The black vertical bars show the statistical uncertainty, and
      the yellow bands show the total systematic uncertainty, except
      for the $5.8\%$ uncertainty on the luminosity.
      The right panels show the data-to-theory ratio with respect to \alpgenpythia{} and \mcfm{} predictions,
      with the blue dashed bands showing the scale-variation uncertainty of each prediction, which is associated with the 
      variation of the renormalization and factorization scales $\mu$ or to
      the combined variation of $\alpha_{s}^{\textrm{CKKW}}$ and $\Lambda_{QCD}$.
      \label{fig:CC_Zjj_Theta_AM}}
  \end{center}
\end{figure*}

Figures~\ref{fig:CC_Pt2_Lead} to~\ref{fig:CC_Zjj_Theta_AM} show
measured differential cross sections in the \Ztwojets{} final
state. Figures~\ref{fig:CC_Pt2_Lead} and~\ref{fig:CC_Y2_Incl} show the
measured cross section as a function of the $2\textit{nd}$ leading jet
\pt{} and inclusive jet rapidity compared to \alpgenpythia{} and
\blackhatsherpa{} predictions. Measured distributions are in good
agreement with the theoretical
predictions. Figure~\ref{fig:CC_DJ_Mass_AM} shows the measured cross
section as a function of the dijet mass, $M_{\mathit{jj}}$. The cross
section in the first bin is overestimated by the \mcfm{} prediction,
but correctly described by the \alpgenpythia{} prediction. In the
$M_{\mathit{jj}}$ region above approximately 160~\gevcsq{}, the
measured cross sections are $10\%-20\%$ higher than both predictions. However, the systematic
uncertainty, mainly due to the jet-energy scale, is as large as the observed difference.
Figure~\ref{fig:CC_DJ_DR_AM} shows the measured cross section as a
function of the dijet $\Delta{R}$ compared to \alpgenpythia{} and
\mcfm{} predictions. Some differences between data and theory are
observed at high $\Delta{R}$, where the measured cross section is
approximately $50\%$ higher than the theoretical predictions. The dijet
$\Delta{\phi}$ and $\Delta{y}$ differential cross sections also are
measured, and the results are shown in Figs.~\ref{fig:CC_DJ_DPhi_AM}
and~\ref{fig:CC_DJ_DY_AM}. The dijet $\Delta{\phi}$ appears reasonably
modeled by the \alpgenpythia{} and \mcfm{} predictions, whereas the
dijet $\Delta{y}$ shows a shape difference, which reaches $50\%$
at $\Delta{y} = 3-3.6$, and is related to the observed difference between
data and theory at $\Delta{R} \gtrsim 4$. This region is affected by
large experimental uncertainties, mainly due to the pile-up subtraction,
and large theoretical uncertainty. Figure~\ref{fig:CC_Zjj_Theta_AM}
shows the measured cross section as a function of the dihedral angle
$\theta_{Z,{\mathit{jj}}}$ between the \Zll{} decay plane and the jet-jet
plane~\footnote{$\theta_{Z,{\mathit{jj}}}$ is defined as $\theta_{Z,{\mathit{jj}}} =
  \arccos{ \frac{(\vec{\ell_1} \times \vec{\ell_2}) \cdot (\vec{j_1} \times
      \vec{j_2})} {|\vec{\ell_1} \times \vec{\ell_2}||\vec{j_1} \times
      \vec{j_2}|}}$, where $\vec{\ell}$ and $\vec{j}$ are the momentum
  three-vectors of leptons and jets.}. The measured cross section is in good
agreement with the \alpgenpythia{} and \mcfm{} predictions.

\subsection{Cross section for the production of a \Zg{} boson in
  association with three or more jets \label{sec:Z3jet_results}}
\begin{figure*}
  \begin{center}
    \subfigure[]{\includegraphics[width=\figsize]{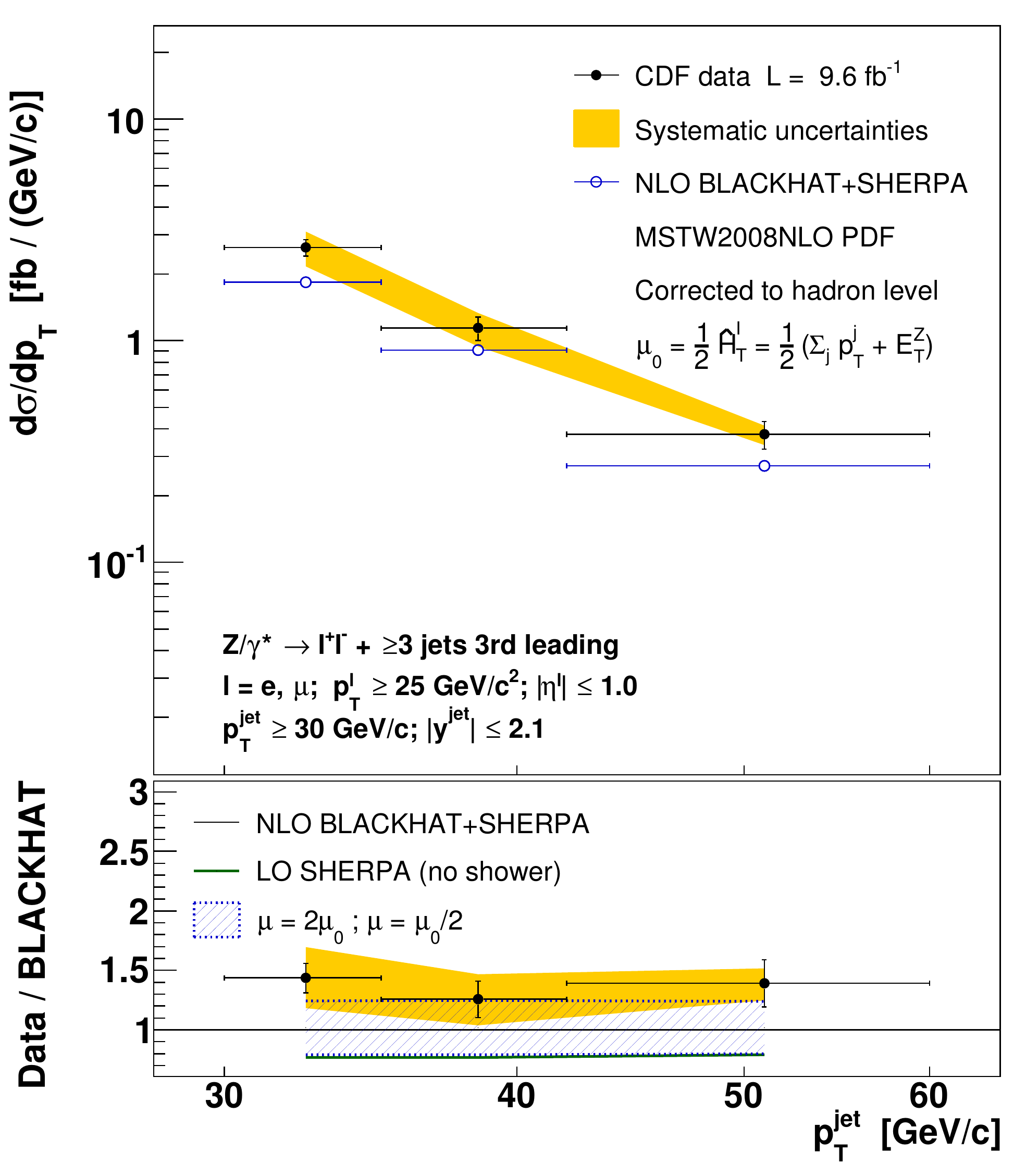}}
    \subfigure[]{\includegraphics[width=\figsize]{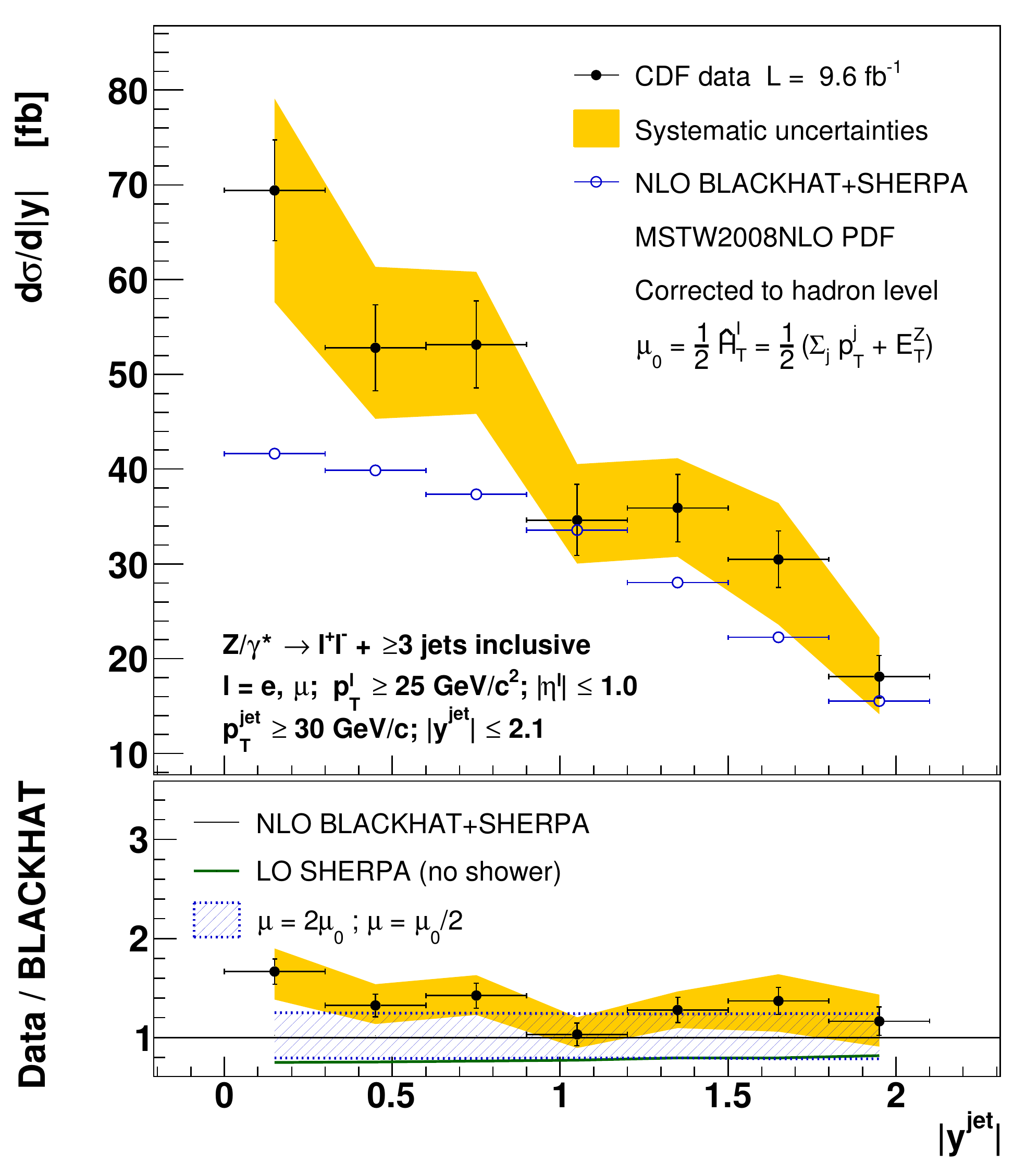}}
    \caption{Differential cross section as a function
      of (a) $3\textit{rd}$ leading jet \pt{} and (b) inclusive jet
      rapidity for \Zthreejets{} events.
      The measured cross section (black dots) is compared to the \blackhatsherpa{} NLO prediction (open circles).
      The black vertical bars show the statistical uncertainty, and
      the yellow bands show the total systematic uncertainty, except
      for the $5.8\%$ uncertainty on the luminosity.
      The lower panels show the data-to-theory ratio,
      with the blue dashed bands showing the scale-variation uncertainty, which is associated with the
      variation of the renormalization and factorization scales $\mu$.
      \label{fig:CC_Pt3_Y3_BH}}
  \end{center}
\end{figure*}

Figure~\ref{fig:CC_Pt3_Y3_BH} shows the differential cross sections as
a functions of $3\textit{rd}$ leading jet \pt{} and inclusive jet
rapidity in events with a reconstructed \Zll{} decay and at least three
jets. The NLO \blackhatsherpa{} prediction is
approximately $30\%$ lower than the measured cross sections for \Zg{} +
$\geqslant 3$ jets events, but data and predictions are still
compatible within the 
approximately $25\%$ scale-variation uncertainty and
the $15\%$ systematic uncertainty, dominated by the jet-energy scale.
Apart from the difference in the normalization, the shape of the measured
differential cross sections is in good agreement with the NLO
\blackhatsherpa{} prediction.


\section{\label{sec:conclusion}Summary and Conclusions}
The analysis of the full proton-antiproton collisions sample collected with the CDF II detector
in Run II of the Tevatron, corresponding to $9.6$~fb$^{-1}$
integrated luminosity, allows for precise measurements of \Zjets{} inclusive and
differential cross sections, which constitute an important legacy of
the Tevatron physics program.
The cross sections are measured using the decay
channels \Zee{} and \Zmm{} in the kinematic region \mbox{$\pt^{\ell}
  \geqslant 25$~\gevc{}}, \mbox{$|\eta^{\ell}| \leqslant 1$},
\mbox{$66 \leqslant M_{\ell^{+}\ell^{-}} \leqslant 116$~\gevcsq{}},
\mbox{$\ptjet \geqslant 30$~\gevc{}}, \mbox{$|\yjet| \leqslant 2.1$},
and \mbox{$\Delta{R}_{\ell-\textrm{jet}} \geqslant 0.7$}, with jets
reconstructed using the midpoint algorithm in a radius $R=0.7$. The
measured cross sections are unfolded to the particle level and the
decay channels combined. Results are compared with the most recent theoretical
predictions, which properly model the measured differential cross
sections in \mbox{$\Zg + \geqslant 1$}, 2, and 3 jets final states. The main
experimental uncertainty is related to the jet-energy scale, whereas the
largest uncertainty of the theoretical predictions is generally
associated with the variation of the renormalization and factorization scales.
Among perturbative QCD predictions, \loopsimmcfm{} shows the lowest
scale-variation uncertainty and, therefore, gives the most accurate
cross-section prediction for the \Zonejet{} final state. The \mcfm{} and
\blackhatsherpa{} fixed-order NLO predictions are in reasonable
agreement with the data in the \mbox{$\Zg + \geqslant 1$}, 2, and 3 jets final
states. The \alpgenpythia{} prediction provides a good modeling of
differential distributions for all jets multiplicities. The \powhegpythia{} prediction, due to
the NLO accuracy of the matrix elements and to the inclusion of
nonperturbative QCD effects, provides precise modeling of \Zonejet{}
final states both in the low- and high-\pt{} kinematic regions. The
effect of NLO electroweak virtual corrections to the \Zg{} + jet
production is studied and included in the comparison with the
measured cross sections: in the high \pt{} kinematic region,
corrections are of the order of $5\%$, which is comparable with the
accuracy of predictions at higher order than NLO.
The large theoretical uncertainty associated with the variation of the
renormalization and factorization scales suggests that the inclusion
of higher order QCD corrections, by mean of exact or approximate
calculations, will improve the theoretical modeling
of \Zjets{} processes.

The understanding of associated production of vector bosons
and jets is fundamental in searches for non-SM physics, and the
results presented in this paper support the modeling of \Zjets{} currently
employed in Higgs-boson measurements and searches for physics beyond the standard model.

\begin{acknowledgments}
We thank the Fermilab staff and the technical staffs of the
participating institutions for their vital contributions. This work
was supported by the U.S. Department of Energy and National Science
Foundation; the Italian Istituto Nazionale di Fisica Nucleare; the
Ministry of Education, Culture, Sports, Science and Technology of
Japan; the Natural Sciences and Engineering Research Council of
Canada; the National Science Council of the Republic of China; the
Swiss National Science Foundation; the A.P. Sloan Foundation; the
Bundesministerium f\"ur Bildung und Forschung, Germany; the Korean
World Class University Program, the National Research Foundation of
Korea; the Science and Technology Facilities Council and the Royal
Society, UK; the Russian Foundation for Basic Research; the Ministerio
de Ciencia e Innovaci\'{o}n, and Programa Consolider-Ingenio 2010,
Spain; the Slovak R\&D Agency; the Academy of Finland; the Australian
Research Council (ARC); and the EU community Marie Curie Fellowship
contract 302103.
\end{acknowledgments}

\cleardoublepage
\bibliography{prdzjets}
\end{document}